\documentclass[twocolumn,showpacs,preprintnumbers,amsmath,amssymb,pre]{revtex4}

\usepackage{graphicx}
\usepackage{dcolumn}
\usepackage{overpic}
\usepackage{float}

\begin{document}

\title{Excited-state quantum phase transitions and periodic dynamics} 

\author{G. Engelhardt}
\email{georg@itp.tu-berlin.de}

\author{V. M. Bastidas}

\author{W. Kopylov}

\author{T. Brandes}

\affiliation{%
Institut f\"ur Theoretische Physik, Technische Universit\"at Berlin, Hardenbergstr. 36, 10623 Berlin, Germany}%

\begin{abstract}
We investigate signatures of the excited-state quantum phase transition in the periodic dynamics of the Lipkin-Meshkov-Glick model and 
the Tavis-Cummings model. In the thermodynamic limit, expectation values of observables in eigenstates of the system can be 
calculated using classical trajectories. Motivated by this, we suggest a method based on the time evolution of the 
finite-size system,  to find  singularities in observables, which arise due to the excited-state quantum phase transition.
\end{abstract}


\maketitle

\section{Introduction}
\label{sec:0}
 Besides the well-known quantum bifurcation in the ground state, known from the Dicke model~\cite{Emary2003} and the bosonic Josephson 
junction~\cite{Chuchem2010}, quantum criticality also takes place in  excited states~\cite{Caprio2008}. 
A key feature of the so-called excited-state quantum phase transitions (ESQPTs),
is a  level clustering at critical energies, which results in a logarithmic singularity in  the density of states (DOS)~\cite{Caprio2008,Ribeiro2008,Brandes2013}.
This singularity is induced by a saddle point of the  semiclassical energy surface~\cite{Caprio2008}. 
Accordingly, in the thermodynamic limit the eigenstates of the quantum system 
also experience symmetry breaking, in the sense of being degenerate or not, depending on their energy~\cite{Puebla2013}.

As a consequence of the strong relation to the classical dynamics, the underlying classical dynamics drives the ESQPT, which can be
considered to be a quantum manifestation of a separatrix~\cite{Scharf1992}.

Among the quantum signatures of the separatrix~\cite{Aubry1996,Chuchem2010,Nissen2010,Caballero-Benitez2010,Juli'a-D'iaz2010,Juli'a-D'iaz2010a,Kawaguchi2012,
Bernstein1993,Franzosi2000,Franzosi2001,Karkuszewski2002,Buonsante2004,Graefe2007},
intriguing relations to spin squeezing and entanglement~\cite{Juli'a-D'iaz2012,Juli'a-D'iaz2012a,Hennig2012} have been explored.
Previous works study a plethora of dynamical consequences of ESQPTs, i.e., 
their influence on quantum quenches \cite{P'erez-Fern'andez2011a}, the 
relation to chaos~\cite{P'erez-Fern'andez2011}, and decoherence of a qubit coupled to a system possessing an ESQPT~\cite{P'erez-Fern'andez2009}. 
In the context of driven quantum systems, singular behavior resembling ESQPTs appears in the quasienergy 
spectrum of the kicked top~\cite{Bastidas2014}.

Experimental signatures of ESQPTs have been found in molecular 
systems~\cite{Winnewisser2005}, in the diverging period of a spinor Bose-Einstein condensate~\cite{Zhao2014}, and in microwave billiards~\cite{Dietz2013}. 
In pioneering experiments, Oberthaler and collaborators realized the anisotropic Lipkin-Meshkov-Glick (LMG) model by coupling the internal
degrees of freedom of a $^{87}$Rb Bose-Einstein condensate~\cite{Zibold2010,Gross2010,Steel1998}. Additionally, 
the striking experimental observation of  Dicke superradiance in Bose gases loaded in a cavity~\cite{Baumann2010,Baumann2011}, 
opens the possibility to explore the physics of excited states, due to the high degree of control of the system.
Recently, there has been a renewed interest in the experimental investigation of the Tavis-Cummings model, which 
is realized using cavity-assisted Raman transitions \cite{Baden2014}.

Motivated by these experimental highlights, in this paper we explore 
the signatures of the ESQPT that appear in the periodic dynamics of the LMG model and  the Tavis-Cummings (TC) model. 
In doing so, we use the fact that 
expectation values in eigenstates can be calculated by a semiclassical temporal average~\cite{Ribeiro2008,Paul1993}.
In contrast to the LMG model, the classical dynamics of the TC model do not occur on a two-dimensional manifold. Therefore, it is 
interesting to investigate 
if our approach is also applicable for models with regular dynamics and a high-dimensional phase space such as the TC model.

The rest of the paper is organized as follows. In Sec.~\ref{sec:I}, we introduce the models that will be discussed 
along the paper, their corresponding semiclassical limits and the method to calculate time-averaged expectation values.
In Sec.~\ref{sec:II}, we discuss our results for the LMG and TC models. In particular, we perform numerical calculations 
for a finite-size system to compare with the semiclassical results. Finally, in Sec.~\ref{sec:III} we discuss the application
of our method to the LMG model for realistic experimental parameters. The conclusions in Sec.~\ref{sec:III} are followed 
by the Appendix, where we discuss in detail the semiclassical calculations for the Tavis-Cummings model.

\section{Models and Methods}
\label{sec:I}

\subsection{Hamiltonians}
The LMG model describes a set of $N$ two-level systems in a transverse field in the $z$-direction which have anisotropic 
interactions~\cite{Lipkin1965,Meshkov1965,Glick1965}. The Hamiltonian reads 
 \begin{equation}
    H_{\text{LMG}} = -h J_z -\frac 1 N \left(\gamma_x J_x^2 + \gamma_y J_y^2 \right),
    \label{LMGHamiltonian}
 \end{equation}
 where $J_\alpha= \frac 12 \sum_{i}^N \sigma_i^\alpha$ and  $\sigma_i^{\alpha}$, with $\alpha \in \{x,y,z\}$ denote the Pauli
 matrices.
The parameters $\gamma_x$ and $\gamma_y$ describe the interaction strength in the $x$ and $y$ directions, respectively.

 To  get a well-defined thermodynamic limit, we restrict the spin Hilbert space to the  subspace with maximal total angular 
 momentum $j= N/2$ as in Refs.~\cite{Emary2003,Ribeiro2008,Brandes2013}.
This allows  to uniquely describe the atomic Hilbert space by using Dicke states $\left|j,m\right\rangle$, which are the eigenstates of  $J_z$~\cite{Dicke1954}.

  In the experimental realization of the LMG model in Ref.~\cite{Zibold2010,Gross2010} by coupling the internal
degrees of freedom of a  Bose-Einstein condensate, one is naturally 
 restricted to $j=N/2$ in the case of a system consisting of $N$ indistinguishable
 bosonic particles \cite{Steel1998}. 
 The underlying reason  is that the spatial wave functions of all bosons are identical
 due to condensation. This implies that the internal degrees of freedom have to be symmetric under permutation. 
 As the Dicke states with $j=N/2$ satisfy this condition~\cite{Dicke1954}, these are the physically relevant states.

The TC Hamiltonian describes an ensemble of two-level systems with level splitting $\omega_{0}$, which are collectively 
coupled with strength $\lambda$ to a cavity field of frequency $\omega$
\begin{equation}
      \label{TCHamiltonian}
             H_{\text{TC}} = \omega \hat a^\dagger \hat a + \omega_0 J_z + \frac{\lambda}{\sqrt{N}} \left( \hat a J_+ + \hat a^\dagger J_- \right),
\end{equation}
 where $\hat a$ and  $\hat a ^{\dagger}$ are bosonic operators and $J_{\pm}= J_x\pm i J_y$~\cite{Emary2003,Keeling2009,Brandes2013,Tavis1968,Narducci1973}.

The bosonic mode 
is described in terms of Fock states $\left|n \right\rangle$. The TC Hamiltonian commutes with the operator 
$
 \hat{\mathcal{M}} = \hat a^\dagger \hat a + J_z.
$
This allows to restrict the basis of the Hilbert space $\left\{|n\rangle\otimes|j,m\rangle \right\}$ to the symmetry-adapted
basis  $\left\{|M-m\rangle\otimes|j,m\rangle \right\}$, where $M=n+m$ is the eigenvalue of $\hat{\mathcal{M}}$~\cite{Tavis1968}.

\subsection{The semiclassical energy landscapes}
\begin{figure}[t]
  \centering
  \begin{minipage}{0.95\linewidth}
  \begin{overpic}[width=1.\linewidth]{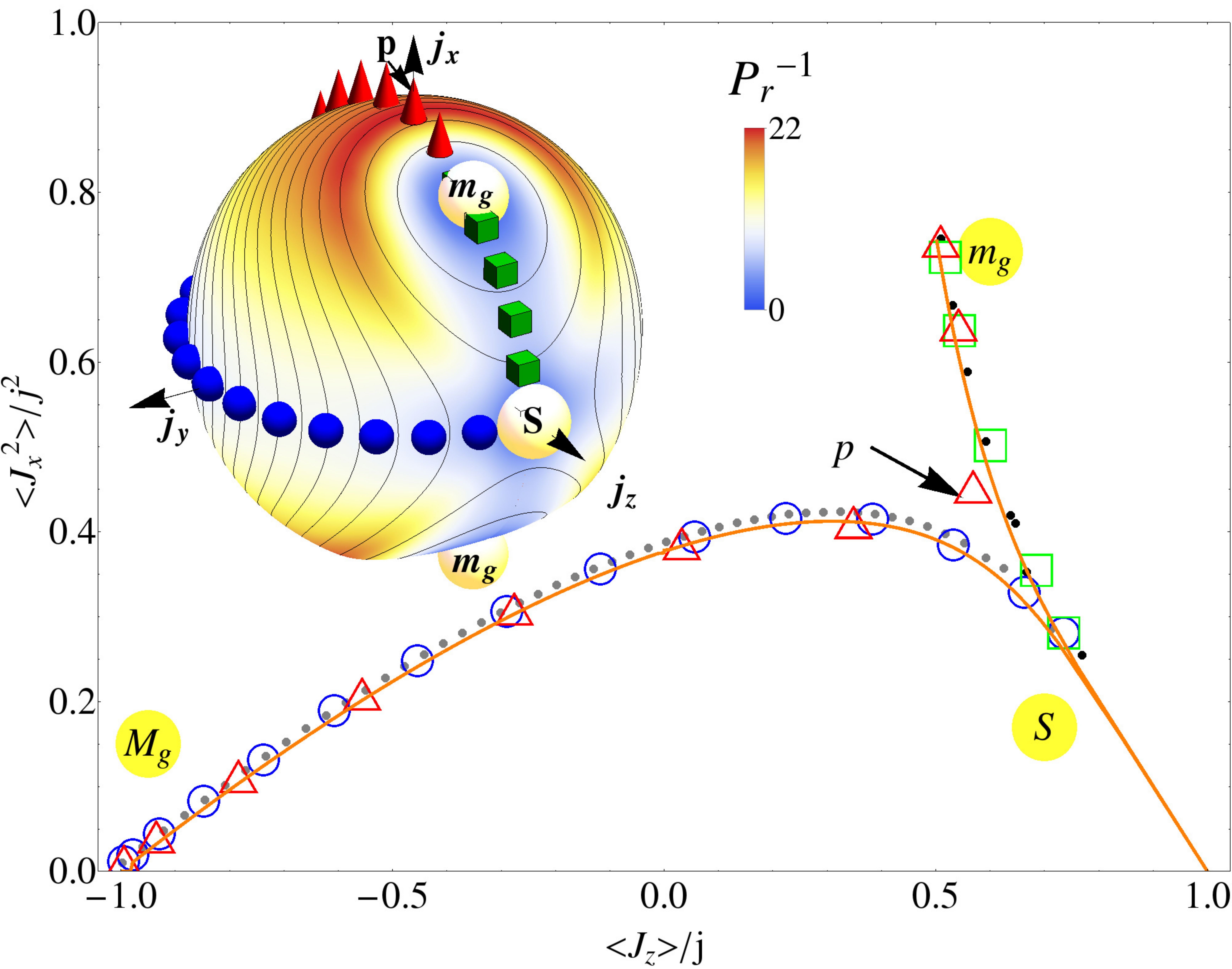}
      \put(-3,75){ \textbf a)}
  \end{overpic}
  \end{minipage}
   \begin{minipage}{0.95\linewidth}
  \begin{overpic}[width=\linewidth]{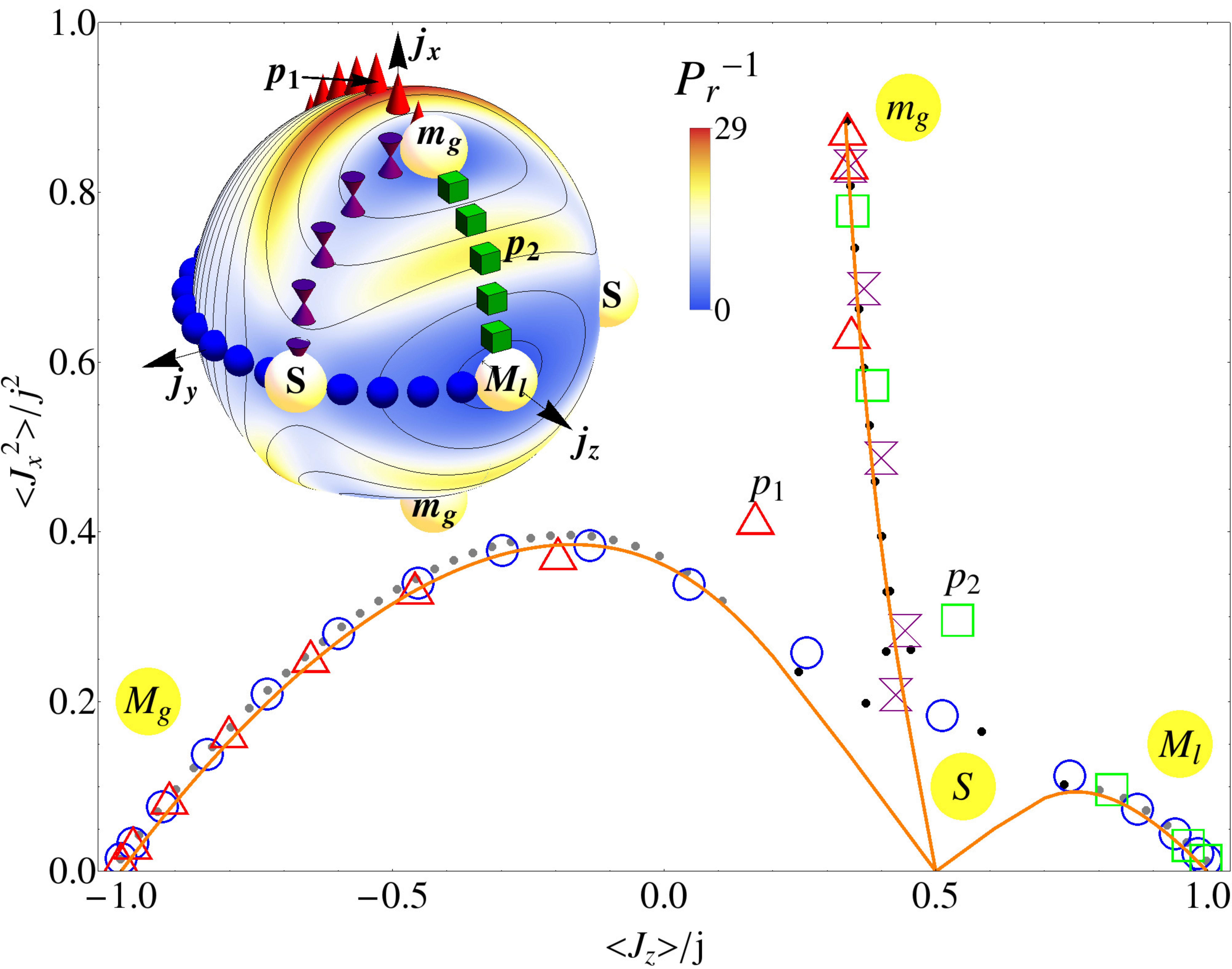}
      \put(-3,75){ \textbf b)}
  \end{overpic}
 \end{minipage}
  \caption{(Color online) 
  (a) Expectation values of $J_z$ and $J_x^2$ for  $\gamma_x/h =2$, $\gamma_y/h=0$  and $j=30$ for the LMG model. 
  (b) Same as in (a) but for $\gamma_x/h =3$ and $\gamma_y/h=2$. For 
  each eigenstate $\left| E_i \right>$  we calculate  $\langle E_i|J_{z}/j|E_i\rangle$ and $\langle E_i|J^{2}_{x}/j^{2}|E_i\rangle$ 
  and place a  dot in the diagram.  Black (gray)  dots correspond  to states with energy less (greater) than the saddle point energy. The solid orange
  line depicts the  semiclassical calculation Eq.~\eqref{eq:ClassicalExp}. The  other symbols (triangles, squares, circles and
  double triangles) correspond to time-averaged expectation values Eq.~\eqref{temporalAverage} for $j=30$ and $\tau = 20/h$.
  The initial conditions are depicted on the Bloch sphere  with the same shape as the symbols in the plot. 
  We mark  states 
  centered at the global minimum, saddle point and  local  maximum with $m_g$,
  $S$ and  $M_l$, respectively, and emphasize the points $p$, $p_1$ and $p_2$. Additionally, we show a density plot of the 
  participation ratio $P_r^{-1}$ on the Bloch sphere.
  The black curves   mark the classical trajectories  with constant energy.  
  }
  \label{fig:ExpectaionValuesC}
\end{figure}

To obtain the semiclassical energy landscape $E_{\text{LMG}}$ of the LMG model,  we scale the Hamiltonian Eq.~\eqref{LMGHamiltonian}
with $j$ and consider the thermodynamic limit $j\gg 1$. 
In this limit, quantum fluctuations are negligible and we can define classical variables $j_{\alpha}=J_{\alpha}/j$ with Poisson 
bracket $\{j_{x},j_{y}\}=j_{z}$ as in Ref.~\cite{Dusuel2005}.
By using these classical variables we define the energy landscape
\begin{equation}
      \label{LMGSemLandscape}
            E_{\text{LMG}}(\textbf{j})=-h j_z -\frac{1}{2} \left(\gamma_x j_x^2 + \gamma_y j_y^2 \right)
            \ ,
\end{equation}
where $\textbf{j}=(j_x,j_y,j_z)$. In this paper we consider the parametrization $\textbf{j}=(\sin\theta \cos\phi,\sin\theta \sin\phi,\cos \theta)$ 
of the Bloch sphere in terms of the local coordinates $(\theta,\phi)$.

In the insets of Fig.~\ref{fig:ExpectaionValuesC}, we depict the level sets of the classical energy $E_{\text{LMG}}$  on the Bloch sphere.
In these figures we mark different kinds of fixed points, namely  local maxima $M_l$, global minima $m_{g}$ and saddle points $S$.
References~\cite{Ribeiro2008,Castanos2006} perform a detailed analysis of the classical energy of the LMG model with respect to the fixed points and present 
a phase diagram of the system. 
Reference~\cite{Ribeiro2008} finds an exact expression for the DOS in the thermodynamic limit, 
and shows that the fixed points  of $E_{\text{LMG}}$ are related to non-analyticities in the DOS. Therefore, a saddle point of the energy 
surface corresponds to a logarithmic singularity and a local maximum to a jump in the DOS.

Due to the continuous symmetry $\hat{\mathcal{M}}$ of the  TC model Eq.~\eqref{TCHamiltonian}, we can derive an effective
semiclassical energy landscape $E_{\text{TC}}$ for the atomic ensemble.

Like for the LMG model, to obtain the semiclassical energy landscape, we scale the Hamiltonian Eq.~\eqref{TCHamiltonian} with $j$. This allows us to use 
the classical variables $\textbf{j}=(j_x,j_y,j_z)$ for the atomic ensemble as in the LMG model. Furthermore, we use the classical variables $a, \vartheta \in \mathbb{R}$ for the bosonic mode
defined by $a e^{i\vartheta}=\hat a /\sqrt{j}$. In addition, we define the classical conserved quantity 
$\mathcal{M}=\hat{\mathcal M}/j=j_z+ a^2$, which enables us to obtain the semiclassical energy landscape 
\begin{align}
         \label{TCRedSemLandscape}
 E_{\text{TC}}(\vartheta,\textbf{j})&=  \omega \mathcal{M} + \left(\omega_0 - \omega \right) j_{z}  
 \nonumber \\&
 + \lambda \sqrt{2\left(\mathcal{M}- j_{z}  \right)}(j_x \cos \vartheta-j_y\sin \vartheta)
 \ .
\end{align}

Finally, we introduce  rotated spin variables  $\tilde j_x = j_x \cos \vartheta - j_y \sin \vartheta$, 
$\tilde j_y = j_x \sin \vartheta + j_y \cos  \vartheta$ and $\tilde j_z = j_z$. In terms of the new coordinates 
$\tilde{\textbf{j}}=(\tilde{j}_x,\tilde{j}_y,\tilde{j}_z)$, the energy landscape
\begin{equation}
         \label{TCRedSemLandscapeRot}
 E_{\text{TC}}( \tilde{\textbf{ j}})=  \omega \mathcal{M} + \left(\omega_0 - \omega \right) \tilde j_{z}  
 + \lambda \sqrt{2\left(\mathcal{M}- \tilde j_{z}  \right)}  \tilde j_x
\end{equation}
represents an effective energy landscape for the atomic ensemble.
In the Appendix we show that the equations of motion of the rotated spins are given by 
$\frac d{dt} \tilde  j_\alpha = \left\lbrace \tilde j_\alpha, E_{\text{TC}} ( \tilde{\textbf{ j}})  \right\rbrace$. The time evolution of 
the bosonic variables follows
\begin{equation}
   a^2 = \mathcal {M} -  \tilde j_z   \qquad  \qquad  \frac d {dt} \vartheta = -\omega -\frac \lambda {a\sqrt 2}  \tilde j_{x}.
\end{equation}
Consequently, the time evolution of $a$,$\vartheta$ depends  on the rotated spins but not the other way around.

\begin{figure}[t]
    \begin{overpic}[width=1.\linewidth]{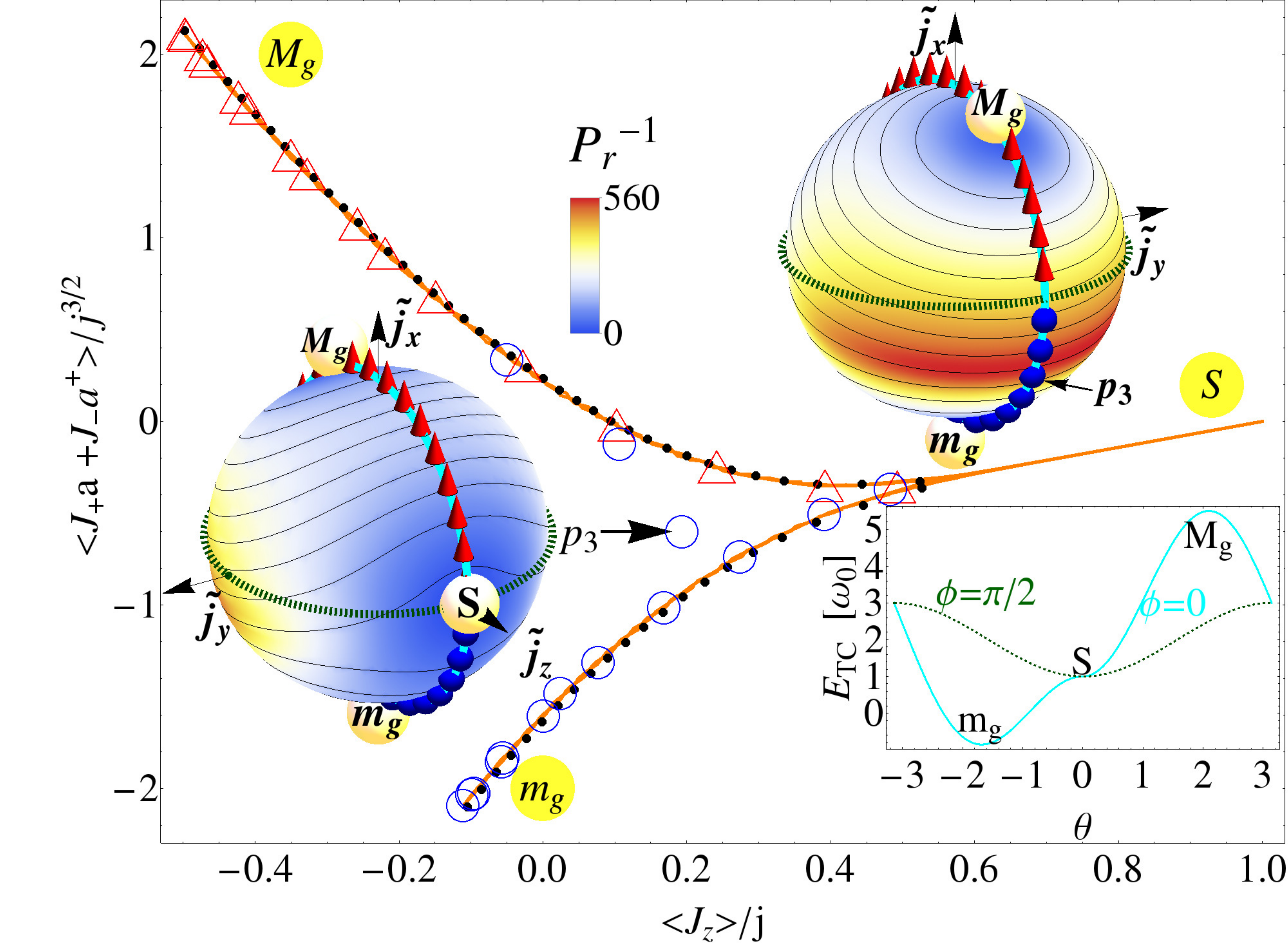}
  \end{overpic}
   \caption{(Color online) 
   Expectation values in eigenstates (black dots)  
   for the TC model~\eqref{TCHamiltonian} for the parameters $\omega / \omega_0=\lambda/\omega_0=2$, $j=30$ 
   and $ M= 30$.
   Time-averaged expectation values~\eqref{temporalAverage} depicted as circles and triangles fulfill $\left< \hat {\mathcal M}\right>=30$.
    The solid orange
  line depicts the  semiclassical calculation~\eqref{eq:ClassicalExp}.
   On the Bloch spheres the black lines depict the level sets of the energy landscape $E_{\text{TC}}\left(\tilde {\mathbf{j}} \right)$. 
   The coloring shows the participation ratio $P_r^{-1}$ of the product of coherent states.
   The triangles and balls mark the positions of the initial states of our simulations.
   The light blue and 
  green curve on the Bloch sphere  mark the sections $\phi=0$ and $\phi=\pi/2$, respectively. In the lower right inset
  we depict $E_{\text{TC}}$  as a function of the local coordinates of the Bloch sphere $(\theta,\phi)$. 
  In particular, we depict the dependence of the energy as a function of
  the polar angle $\theta$ for the azimuthal angles $\phi=0$ and 
  $\phi=\pi/2$. Here one can clearly see the saddle point
   at $\theta=0$.}
  \label{fig:JcPlot}
\end{figure}

In the inset of Fig.~\ref{fig:JcPlot} we depict the level sets of the energy landscape $E_{\text{TC}}$ on the Bloch sphere. 
For $\mathcal M =1$, the energy surface  $E_{\text{TC}}( \tilde{ \textbf{ j}})$ possesses a saddle point, giving rise to an 
ESQPT~\cite{Narducci1973,Bastarrachea-Magnani2014a,P'erez-Fern'andez2011}.

\subsection{Time-averaged expectation values}
According to Refs.~\cite{Ribeiro2008,Paul1993,Ribeiro2009},
the expectation value $\langle\hat{O}\rangle_{E}$ of an operator $\hat O$ in an eigenstate of the system can be related
to the dynamics of the corresponding classical observable $o(t)$. At leading order in $1/j$, such a relation reads 
 \begin{equation}
        \label{eq:ClassicalExp}
        \langle\hat{O}\rangle_{E} = \frac 1 L \sum_{l=1}^L \frac 1 {T_l(E)} \int_0^{T_l(E)} o^{(l)}(t) dt  
       \ ,
 \end{equation}
where $L$ denotes the number of connected trajectories defined by the relations $E=j E_{\text{LMG}}(\textbf{j})$ for the LMG 
model and $E=j E_{\text{TC}}(\vartheta,\textbf{j})$ for the TC model. In Eq.~\eqref{eq:ClassicalExp} we consider the representation
$o^{(l)}(t)$ of the observable $o(t)$ restricted 
to the $l$-th trajectory with period $T_l(E)$. Furthermore, one has to sum over the trajectories $l$ in such a way that their union
has  the symmetry of the  underlying system, i.e., the symmetries of Hamiltonians~\eqref{LMGHamiltonian} and~\eqref{TCHamiltonian}.

References~\cite{Brandes2013,Ribeiro2008} show how to express the expectation values of observables in eigenstates in terms of the DOS. 
As a consequence, the expectation values inherit the singularities of the DOS.

Based on Eq.~\eqref{eq:ClassicalExp}, we now suggest a method which could be used to experimentally detect a signature of the ESQPT.  
The measurement of the energy of the system is often not 
experimentally accessible, which motivates us to employ an alternative representation of the observables.  In order to resolve the
singularities of measurable quantities, we choose two observables $\hat O_1$ and $\hat O_2$. In the numerical simulations we
take $(\hat O_1,\hat O_2)= \left(\frac{J_z}{j},\frac{J_x^2}{j^{2}}\right)$ for the LMG model and
$(\hat O_1,\hat O_2)=\left(\frac{J_z}{j},\frac{\hat a J_+ + \hat a^\dagger J_-}{j^{3/2}}\right)$ for the TC model.

In contrast to the LMG model, the classical dynamics of the TC model  is not restricted to a two-dimensional manifold.
For this reason, we discuss the applicability
of our method to this more complicated model below.

As spin-coherent states $|\theta,\phi\rangle$
are  the  closest ones to classical states~\cite{Arecchi1972}, we take 
these as initial conditions for our quantum-mechanical simulations of the finite-size LMG model. 
A spin-coherent state is obtained by a rotation of the Dicke-State $\left| j, j \right>$ so that  its mean is  located at the 
Bloch sphere coordinates $ (\theta,\phi) $.
More precisely, one can show that  $|\theta,\phi\rangle=  (1+\left|\rho \right|)^{-j}e^{\rho J_-}\left| j, j \right>$,
where $\rho= e^{i \phi} \tan \frac\theta 2$.
In the basis of Dicke states,  a spin-coherent state reads 
$\sum_{m=-j}^{j} t_m    \left| j,m \right> $, where \cite{Zhang1990}
\begin{equation}
 t_m = \sqrt{\left( \begin{array}{c}
		  2j \\
		  j+m 
		  \end{array}   \right) }  \left(\sin\frac{\theta}{2} \right)^{j-m}  \left(\cos\frac{\theta}{2 } \right) ^{j+m} e^{-i (j+m) \phi}.
		  \label{eq:coherentStateCoeff}
\end{equation}

For the TC model, one can use a product  of coherent states of both the spin system as well as the bosonic mode 
(see below for  additional information).

We initially prepare the system in a (product of) coherent state(s) $\left| \psi(0) \right>$.
Afterwards, one measures the expectation values $\langle\hat O_1\rangle$ and $\langle\hat O_2\rangle$ in the state $\left| \psi(t) \right>$,
which allows us to define the temporal average 
 \begin{equation}
 \overline{\left< O_i\right>} = \frac 1 \tau \int_0^{\tau} \left<\psi(t)\right|\hat O_i \left|\psi(t) \right>   dt     ,
  \label{temporalAverage}
\end{equation}
where $\tau$ is the evolution time.
This sequence is repeated for a set of different initial states. 
For a notational reason we  define  $\boldsymbol \chi  =\left(\left< O_1\right>, \left< O_2 \right>\right)$. For the LMG model we 
specify $\boldsymbol \chi_{\left(\theta,\phi \right)}   $ at which the expectation values correspond to Eq.~\eqref{temporalAverage}
with the initial state $\left| \theta,\phi \right>$.

\section{Results}
\label{sec:II}
From an experimental point of view,  an energy-independent representation of the expectation values has the big advantage that it is not necessary to measure the 
energy. Furthermore, such a representation would be useful to study systems in which energy is 
not a conserved quantity like in dissipative~\cite{Morrison2008a,Kopylov2013} or in feed-back systems~\cite{Kopylov2015}. 
Thus, in doing so one can examine the implications of 
an ESQPT under nonequilibrium conditions. 

\subsection{LMG model}

In Fig.~\ref{fig:ExpectaionValuesC} we compare  the expectation values of observables in eigenstates,
the semiclassical calculation found by using Eq.~\eqref{eq:ClassicalExp}, and the quantum-mechanical
averaging method Eq.~\eqref{temporalAverage}. 
In the thermodynamic limit, the expectation values $\left< J_z\right>$ and $\left< J_x^2\right>$ calculated  using 
Eq.~\eqref{eq:ClassicalExp} describe a parametric curve as a function of 
energy, which we denote with $\boldsymbol \chi_{cl} (E)$. Accordingly, we denote the expectation values in eigenstates 
with $\boldsymbol \chi_{es} (E)$.
In Fig.~\ref{fig:ExpectaionValuesC}~(a) there is a cusp of $\boldsymbol \chi_{cl}(E)$ at $\boldsymbol \chi  =(1,0) $.  
This is a result of the ESQPT~\cite{Ribeiro2008}, as both expectation  values exhibit  a singular behavior there in 
the thermodynamic limit as a function of energy. 
The critical energy $E_{S}$ corresponds to the energy of the separatrix, which is  $E_S=j E_{\text{LMG}}(\textbf{j})$ in 
Fig.~\ref{fig:ExpectaionValuesC}~(a).

The expectation values of observables in eigenstates approximately agree with the semiclassical calculation.
However, for finite sizes the expectation values in eigenstates are not directly located at $\boldsymbol \chi =(1,0) $,
because these points can  be achieved only in the thermodynamic limit.

In Fig.~\ref{fig:ExpectaionValuesC}(b) the curve  $\boldsymbol \chi_{cl} (E)$ representing the 
semiclassical calculation exhibits a qualitatively 
different behavior.
In contrast to Fig.~\ref{fig:ExpectaionValuesC}(a) there is no cusp at the saddle point. 
Starting from the global minimum $m_g$ and  increasing the energy $E$, the curve $\boldsymbol \chi _{cl} (E)$ 
exhibits a bifurcation at the 
saddle point energy $E_S$. One branch continues to expectation values corresponding to the global maximum $M_g$ and the 
other one continues to expectation values corresponding to  the local maximum $M_l$. Thus, this 
bifurcation is due to  the 
emergence of a  local maximum of the energy surface~\cite{Ribeiro2008} which can be seen in the inset.

Most of the  expectation values in eigenstates
 are well described by the classical calculation, but there are significant deviations for states close to the saddle point.
 States with energies less than the saddle point energy are nearly degenerate. Therefore, two states that are nearly degenerate have similar 
 expectation values for the chosen observables. We color these expectation values black in Fig.~\ref{fig:ExpectaionValuesC}
, while  the others are depicted in gray. 

For the simulations of Eq.~\eqref{temporalAverage} we use the
initial states depicted on the Bloch spheres in the insets of Fig.~\ref{fig:ExpectaionValuesC}(a) and (b).  These are the
most interesting initial states containing all fixed points of the energy landscape Eq.~\eqref{LMGSemLandscape}. 
Therefore,  trajectories close to every possible energy in the system can be probed \cite{Bastidas2014}.

In Fig.~\ref{fig:ExpectaionValuesC}(a), 
the finite-size simulation agrees with the semiclassical and 
eigenstate calculations. 
The curve $\boldsymbol \chi_{\left(\theta,\phi\right)}$ exhibits a cusp close to $\boldsymbol \chi  = (1,0)$, although the 
energy of the spin-coherent state is smooth as a function of $\left( \theta,\phi  \right)$. For this reason
the path $\boldsymbol \chi_{\left(\theta,\phi \right)}$ shows a signature of the non-analytic character of the ESQPT.

There is  no point exactly located  at the  cusp at $\boldsymbol \chi  = (1,0)$, although the system is
initialized  at the saddle point.
This  is a consequence of the quantum-mechanical deformation of the wave packet being prepared in the vicinity of the saddle point. 
The enhanced quantum fluctuations due to the influence of the saddle point cause a collapse and revival behavior in the time evolution
of observables as it is discussed in Ref.~\cite{Tonel2005}. This also contributes to the  deviation from the semiclassical calculation visible
in Fig.~\ref{fig:ExpectaionValuesC}.

The deformation of the wave packet has been investigated experimentally in Refs.~\cite{Zibold2010,Gross2010,Gerving2012}.
In contrast, wave packets
remaining essentially Gaussian  resemble the semiclassical calculation much more. Detailed investigations of the deviations of semiclassical and quantum dynamics
 can be found in Refs.~\cite{Nissen2010,Chuchem2010,Caballero-Benitez2010,Hennig2012,Buchleitner2002}.
 This argumentation also applies to Fig.~\ref{fig:ExpectaionValuesC}(b), where the finite-size simulation is unable to resolve the bifurcation point 
 $\boldsymbol \chi  = (1/2,0)$.

The initial states located at points $p$, $p_1$ and $p_2$  marked in the insets of Fig.~\ref{fig:ExpectaionValuesC}  lying close to the separatrix also
experience  a strong deformation. Consequently, they considerably deviate from the semiclassical limit.

Deviations of the quantum-mechanical calculation 
from the semiclassical results are strongly related to the  participation ratio $P_r^{-1}$ of the initial
states $\left| \psi(0) \right>=\left|\theta, \phi \right>$,   with $\left|\theta, \phi \right>$ being a spin-coherent state. 
Following Refs.~\cite{Scharf1992,Weaire1977}, we define the ``inverse participation ratio'' as
$
 P_r=  \sum_{i}  \left| \left<\psi(0) \right. \left| E_i\right> \right|^4,
$
where $\left| E_i \right>$  denotes an eigenstate of the system. 
The  participation ratio is  a measure of the number of eigenstates needed to construct a specific state. Therefore, a high 
participation ratio means that our initial state is a superposition of a lot of eigenstates.

In the  insets of Fig.~\ref{fig:ExpectaionValuesC}  we depict a density plot of $P_r^{-1}$ on the Bloch sphere. 
In particular, the initial points
$p$, $p_1$ and $p_2$ in Fig.~\ref{fig:ExpectaionValuesC} which exhibit quite a strong deviation from the semiclassical calculation, 
have a high  participation ratio. 
Consequently, the temporal average of these initial states is influenced by many eigenstates, so that it deviates strongly from 
the semiclassical calculation. The relation between the participation ratio and deviations of quantum dynamics is addressed in Refs. 
\cite{Chuchem2010,Hennig2012,Khripkov2013,Caballero-Benitez2010}.
To improve the simulation close to $\boldsymbol{\chi}=(1/2,0)$ in Fig.~\ref{fig:ExpectaionValuesC}(b), we suggest new initial conditions, which we depict with purple 
double triangles  in the inset.
We choose therefore the points, at which the classical velocity is minimal~\cite{Berry1972} which possess a very low  
participation ratio.

The time-averaged expectation values $\overline{\left< O_i\right>}_{(\theta,\phi)}$ converge to the corresponding
semiclassical ones $\langle\hat{O}_{i}\rangle$ of 
Eq.~\eqref{eq:ClassicalExp}. 
Given a finite size $j$, the scaling 
$|\langle\hat{O}_{i}\rangle-\overline{\left< O_i\right>}_{(\theta,\phi)}| \propto 1/\log j$ for an initial
state at a saddle point has been discussed in Ref. \cite{Chuchem2010} for $\gamma_y=0$. We also checked this scaling 
numerically for $\gamma_y\neq 0$ and for other initial states at the separatrix .

 \subsection{TC model}
 
 For the TC model, Fig.~\ref{fig:JcPlot} shows the expectation values of the observables 
$\hat{O}_1=J_{z}/j$ and $\hat{O}_2=(\hat{a} J_{+}+ \hat{a}^{\dagger} J_{-})/j^{3/2}$ using the different calculation techniques.  
 As explained before,  the classical energy surface exhibits a saddle point for $M=j$.
For this reason, black dots depict  the observables in eigenstates $\boldsymbol \chi_{es} (E)$ for a finite-size system with $j = M =30$.
Although the classical dynamics is not restricted to a two-dimensional manifold for the TC model, the expectation values
in eigenstates can be calculated semiclassically with a high accuracy. The  ESQPT cusp is located at $\boldsymbol \chi = (1,0)$. However, as for the 
LMG model, $\boldsymbol \chi_{es} (E)$  reaches this point only in the thermodynamic limit.

As stated before we suggest a product of spin and bosonic  coherent states
\begin{equation}\label{TC_coh_state}
  \left|\psi(0)\right> =  \left|\alpha \right>\otimes\left|\theta,\phi\right>
\end{equation}
as initial state, where $\left| \alpha \right>$ denotes a coherent state of the bosonic mode~\cite{Glauber1963}.
Its mean photon number is  given by $\left< \hat a^\dagger \hat  a \right> = \left| \alpha  \right|^2 $. 
In order to satisfy the condition $M=j$, 
the initial state $\left|\psi(0)\right>$ shall fulfill 
\begin{equation}
 \left< \hat{\mathcal M} \right> = j \cos \theta + \left| \alpha  \right|^2 = j.
 \label{eq:initialConstrain}
\end{equation}
This constrain is also fulfilled by the expectation value of the time-evolved state as $\hat {\mathcal M}$ commutes 
with the Hamiltonian. 
Due to its definition,  the initial state~\eqref{TC_coh_state} is not restricted to the subspace $M=j$ and needs the whole Hilbert
space to be defined. 
The variance of $\hat{\mathcal M}$ in the proposed initial state is $\text{Var}\; \hat{\mathcal M} = j \left(1- \cos \theta + \frac 12\sin^2\theta \right)$.
Thus,
in the thermodynamic limit  the variance  of the scaled quantity $\hat{\mathcal M}/j$
scales as  $1/j$.

We can use the symmetry $\hat{\mathcal M}$ to reduce the numerical effort.
To this end, we decompose the initial state in a sum of states with different quantum numbers $M$.
Therefore, we use the representation of spin and bosonic coherent states in terms of Fock and Dicke states, respectively \cite{Zhang1990}.
We write the initial state in Eq.~\eqref{TC_coh_state} 
\begin{equation}
 \left|\psi(0) \right> 
                                           = \sum_{M=M_{min}}^{M_{max}} \sum_{m=-j}^{min(j,M)} a_{M-m} t_m  \left| M-m \right>  \otimes  \left| j,m \right> , \\ 
  \label{eq:prodCoherentStates}
 \end{equation}
where 
$(M_{min}, M_{max})=\left(-j,\infty\right)$,
\begin{align}
a_n &= e^{-\frac{\left|\alpha  \right|^2}{2}} \frac{\alpha^n}{\sqrt{n!}}  
\end{align}
and $t_m$ is given in~\eqref{eq:coherentStateCoeff}
in terms of the symmetry-adapted basis discussed in Sec.~\ref{sec:I}. 
Therefore, $a_n$ and $t_m$ denote the coefficients of the bosonic and spin-coherent states, respectively.
The amplitude of $\alpha$ is fixed due to the constrain~\eqref{eq:initialConstrain}. We choose the phase of $\alpha$ to be
$\vartheta = \arg \alpha = 0 $, thus $\alpha = \sqrt{j-j \cos \theta}$.
The time evolution for different $M$ for that initial state decouples due to the symmetry $\hat{\mathcal M}$, which 
reduces the numerical effort.
In the numerical calculation  we can  truncate the state at
$(M_{min},M_{max}) = (j-\Delta M, j+\Delta M)$ with $\Delta M$ chosen in such a way that the time evolution 
of the expectation values converges.  

Figure~\ref{fig:JcPlot} depicts the results of the  time-averaged quantum simulations. 
 The initial conditions are 
sketched on the Bloch sphere  with blue balls and red triangles  located along the paths
$\theta \in \left(0, \pi\right)$ for $\phi=0$ and $\phi=\pi$, respectively. The time-averaged expectation values
are depicted  with  corresponding symbols.
At this point we recall that the energy
surface on the Bloch sphere is depicted in a rotated frame for which the rotation angle is given by $\vartheta$. As we choose
$\vartheta=0$ for our initial conditions, the rotated frame is equivalent to  the laboratory frame. 

The result of the finite-size simulations resembles the findings
for the LMG model in Fig.~\ref{fig:ExpectaionValuesC} (a). The initial conditions close to the saddle point of the classical
energy surface reproduce the expectation values of the semiclassical calculation with a high accuracy. However,   
for an initial condition located on the separatrix but away from the saddle point, there are also significant
deviations from the semiclassical calculation. In Fig.~\ref{fig:JcPlot} we denote this point with $p_3$. We also
find that this initial condition is characterized by a high participation ratio $P_r^{-1}$, 
which smoothens the signature of the ESQPT.

\section{Applications}
\label{sec:III}
Finally, we discuss the experimental applicability of our method for the LMG model. Here, we refer to the 
 experimental realization  of the LMG model  in Refs.~\cite{Zibold2010,Gross2010}. 
 This experimental realization allows us to prepare the system in a spin-coherent state on arbitrary 
 positions on the Bloch sphere~\cite{Oberthaler}. The measurement of the expectation values of $J_z$ and $J_x^2$ 
 is performed
 by repeating the time evolution for the same initial state up to a given time $t$. Based on the 
 single measurements one can then calculate the desired expectation value at time $t$.
 
\begin{figure}[t]
  \begin{overpic}[width=1.\linewidth]{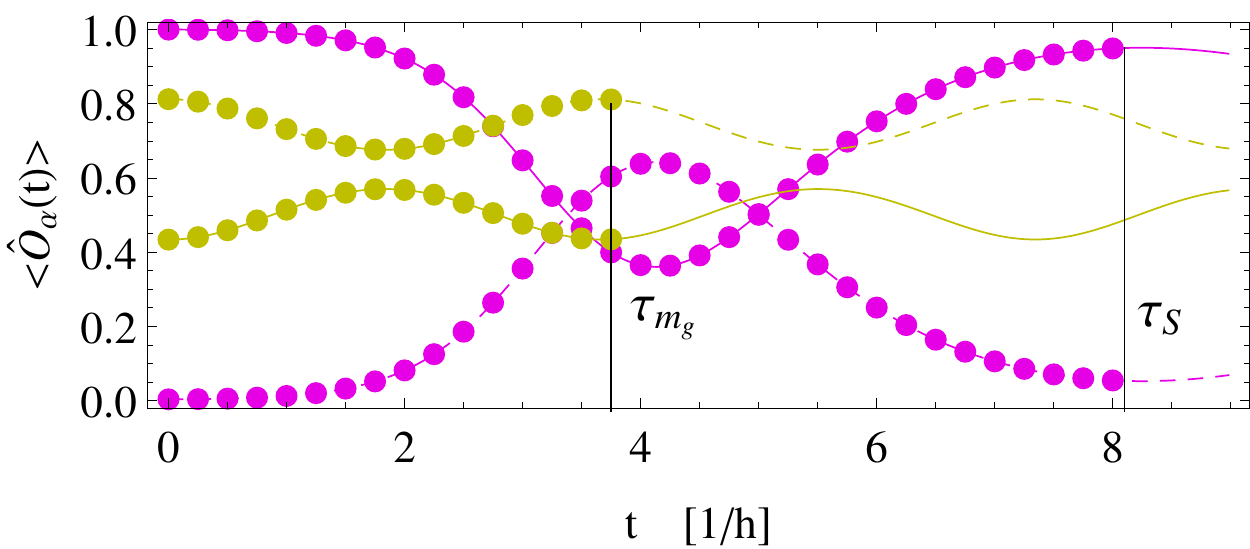}
  \put(-3,40){ \textbf a)}
  \end{overpic}
 \begin{overpic}[width=1.\linewidth]{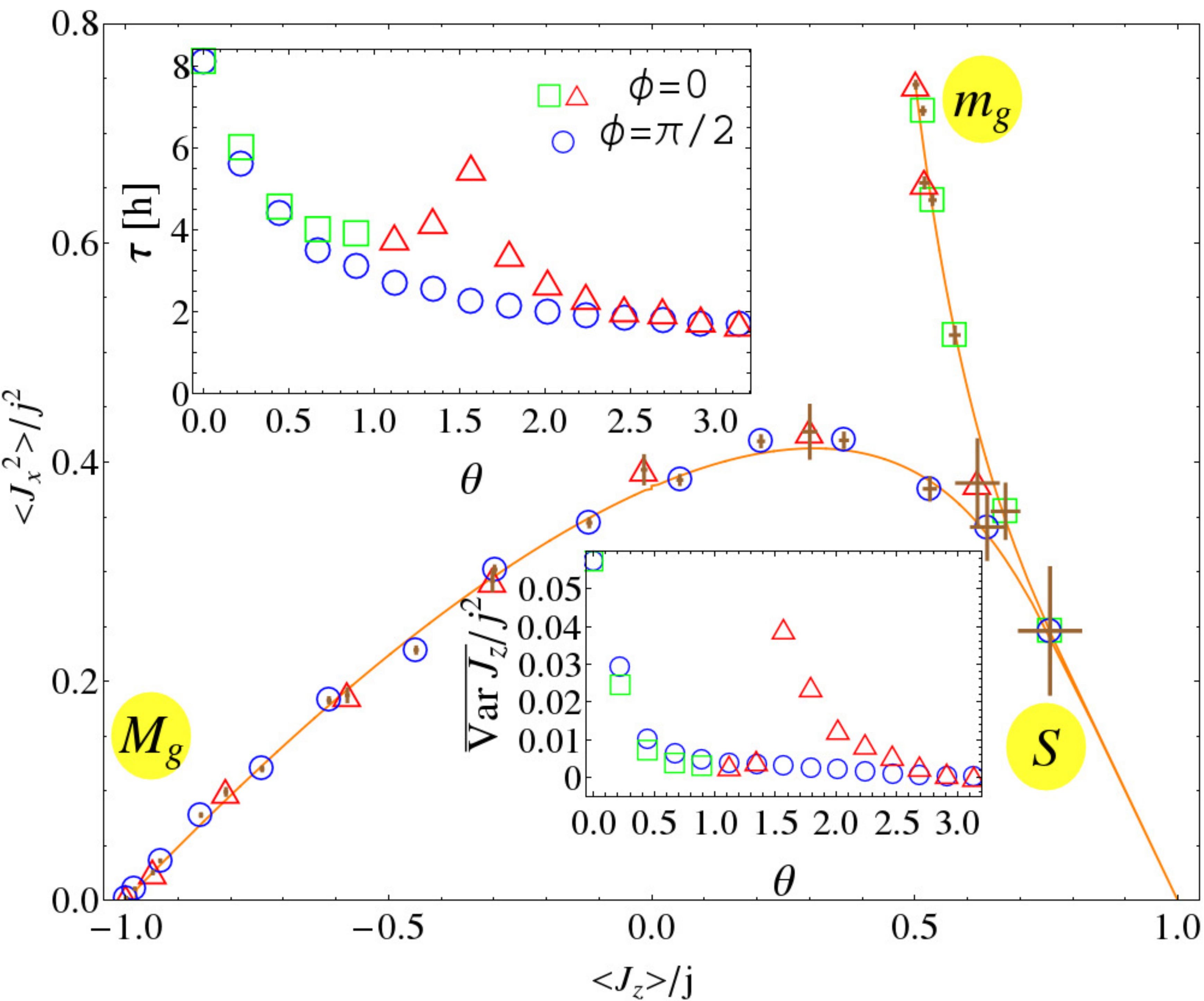}
   \put(-3,75){ \textbf b)}
  \end{overpic}
   \caption{(Color online) 
  (a) Example of the time evolution  of $\hat O_1=J_z/j$ (solid lines) and $\hat O_2=J_x^2/j^2$ (dashed lines) in the LMG model for initial states located at the saddle point [pink (dark gray)]
  and close to a global minimum [yellow (light gray)].   Parameters are the same as in Fig.~\ref{fig:ExpectaionValuesC}(a), but $j=150$.  
   The black lines depict 
  the estimated periods $\tau_{S,m_g}$ for the chosen initial states, respectively.
  The dots mark the times $t_k$ used to evaluate Eq.~\eqref{temporalAverageDiscrete} at which the time step  is chosen to be $\Delta t = \frac 1{4h} $.
  (b) Realistic simulation of the method for the LMG model. The colored symbols and the orange line depict the same as in Fig.~\ref{fig:ExpectaionValuesC}(a).
  The period $\tau$ is 
  always estimated by the time evolution of the observables and depicted in the top inset.  Here we also used $\Delta t = \frac 1{4h} $.
  The bars depict the time-averaged variance of the observables which is defined in \eqref{eq:temporalAverageVar}. 
   In the bottom inset we depict the time-averaged variance of $J_z$. }
   \label{fig:expTest}
\end{figure}

 The maximum feasible time $\tau$ should be long enough to observe recurrences in the time evolution of observables for all initial 
 states~\cite{Oberthaler}.
 In Fig.~\ref{fig:expTest} we simulate these experimental circumstances for an experimentally feasible
 particle number $N=300$ and $\gamma_x/h=2$. Based on the realization of the LMG model in Ref.~\cite{Zibold2010,Gross2010},  experimentally feasible parameters 
 are $h = 2\pi \times 9.45\ \mathrm{Hz}$ and $\gamma_x= 2\pi \times 18.9\ \mathrm{Hz} $.
 From the time evolution of 
 $\left< J_z(t)\right>$ and  $\left< J_x^2(t)\right>$ we estimate the duration
 of one period $\tau$ for each initial preparation. 
 As  examples  we depict in Fig.~\ref{fig:expTest}(a) the time evolution  of an  initial state located at the saddle point and 
 of an initial state located close to the global minimum. We also mark the respective estimated period $\tau$.
 In the inset of Fig.~\ref{fig:expTest}(b) we show the chosen $\tau$ for the initial states 
 depicted in the inset of Fig.~\ref{fig:ExpectaionValuesC}(a). As continuous measurements are not possible, we consider a 
 discretized version of Eq.~\eqref{temporalAverage}, namely
 \begin{equation}
 \overline{\left< O_i\right>}_{\left(\theta,\phi \right)} = \frac 1 \tau \sum_{k=0}^{n_{(\theta,\phi)}-1}  \left<\psi( t_k)\right|\hat O_i \left|\psi( t_k) \right>   \Delta t    ,
  \label{temporalAverageDiscrete}
\end{equation}
where the time step $\Delta t $ and $n_{(\theta,\phi)}$ fulfill    $\Delta t  n_{(\theta,\phi)} = \tau  $
 and $t_k = k  \Delta t$. 
To minimize the experimental effort it will be convenient to take  $\Delta t$  as large as possible.
In Fig.~\ref{fig:expTest} we choose $\Delta t=\frac{1}{4h} $, which still enables a sufficient precision. 
For the  parameters given above this means that $\tau \leq 134\ \mathrm{ms}$ and $\Delta t = 4.2\ \mathrm{ms}$, so that at most
$n_{(\theta,\phi)}=32$ for an initial state at the saddle point.
As a demonstration, in Fig.~\ref{fig:expTest}(a) we   mark the points used in the average~\eqref{temporalAverageDiscrete} with dots. Due to our 
choice of $\Delta t$ these points are dense in relation to the temporal variation of the observables. As the chosen
$\tau$ are quite small, the average does not suffer from the complex collapse and revival behavior appearing for longer 
evolution times observed in Ref.~\cite{Tonel2005}.

As explained before, depending on the initial condition the state can be strongly deformed. This hinders the 
measurement of the expectation values as one has to repeat the measurement more often to obtain the required
precision. To estimate this effort, we also included bars for each point in Fig.~\ref{fig:expTest}(b) depicting
the corresponding time-averaged variance of the time evolution,
 \begin{equation}
  \overline{\text{Var} \;\hat O_i} = \frac 1 \tau \sum_{k=0}^{n_{(\theta,\phi)}-1} \text{Var} \; \hat O_i(t_k)         \Delta t    
  \label{eq:temporalAverageVar}
\end{equation}
where
 \begin{equation}
 \text{Var} \;\hat O_i(t) =  \left<\psi( t )\right|\left(  \hat O_i - \left<\hat O_i(t) \right>  \right)^2 \left|\psi( t) \right>     
  \label{eq:VarO}
\end{equation}
is the variance of the observable $\hat O_i$ at time $t$. 
We  depict the time-averaged variance of $\hat O_i=J_z$ in the inset of  Fig.~\ref{fig:expTest}(b).
We see that for initial states away from the saddle point the variance is vanishing small, as the states
remain essentially Gaussian, but for states close to the separatrix one has to repeat the time evolution for
a given time $t$ quite often.

The time-averaged variance defined in Eq.~\eqref{eq:temporalAverageVar} does  account only for the quantum fluctuations  of the system. 
However, the actual uncertainties appearing in our method depend on the experimental realization.
For example, one could also consider other influences, e.g., uncertainties in preparing the initial
state. However, in the context of the experimental realization of Ref.~\cite{Gross2010,Zibold2010} we assume this to have a
minor influence on the measured results, as there is a high degree of control of the preparation of the initial state.

Referring to Ref.~\cite{Gross2010}, we assume that about $n_{\text{EM}}=60$ experimental measurements are sufficient to determine adequately
the expectation values of $\hat O_i$ for each time step in Fig.~\ref{fig:expTest}(a). Consequently, to obtain the  point in Fig.~\ref{fig:expTest}(b)
corresponding to the saddle point, one has to perform  $n_{\text{SEM}} =2 n_{\text{EM}}  n_{(\theta,\phi)} = 3840$ single experimental measurements,
where $n_{(\theta,\phi)}=32$ for the saddle point as in Eq.~\eqref{temporalAverageDiscrete}. We note that in the experimental realization presented
in Ref.~\cite{Muessel2014} up to $n_{\text{PM}}=30$ measurements  can be performed in parallel in one experimental run so that one
needs $n_{\text{SEM}}/n_{\text{PM}}=128$ experimental runs. For the other points in Fig.~\ref{fig:expTest}(b) fewer experimental runs are necessary.

\section{Conclusions}
\label{sec:IV}
Expectation values of quantum-mechanical operators in eigenstates can be calculated using the classical dynamics of the system. 
This motivated us to
suggest a method which opens a new avenue to experimentally detect signatures of the ESQPT  in systems in which energy is not experimentally accessible or not conserved.

The temporal averaging of a finite-size system resembles the expectation values
in eigenstates and the semiclassical calculations.
However, there are 
partial deviations  from the semiclassical calculations  for energies near an ESQPT. These deviations 
are bigger for initial states obeying a high   participation ratio.

A point to be addressed in the future is the application of our method to more complicated mean-field type models  such as spinor
Bose-Einstein condensates \cite{Hamley2012,Gerving2012}  and the Dicke model \cite{Baumann2010,Baumann2011}. 
In particular, 
our representation of observables might be 
interesting in the description of 
nonequilibrium systems, in which the energy is not a conserved quantity, such as driven~\cite{Bastidas2012b,Engelhardt2013}, 
dissipative~\cite{Morrison2008a,Kopylov2013}  or feedback~\cite{Kopylov2015} mean 
field-type systems.

\begin{acknowledgments}
The authors gratefully acknowledge financial  support from the DFG Grants  BR 1528/7, BR 1528/8, BR 1528/9, SFB 910 and GRK 1558 and  inspiring
 conversations with P. P\'erez-Fern\'andez, M. Vogl and M. Oberthaler and the members of his group.
\end{acknowledgments}

 \bibliography{Bibliography}

\begin{thebibliography}{64}
\expandafter\ifx\csname natexlab\endcsname\relax\def\natexlab#1{#1}\fi
\expandafter\ifx\csname bibnamefont\endcsname\relax
  \def\bibnamefont#1{#1}\fi
\expandafter\ifx\csname bibfnamefont\endcsname\relax
  \def\bibfnamefont#1{#1}\fi
\expandafter\ifx\csname citenamefont\endcsname\relax
  \def\citenamefont#1{#1}\fi
\expandafter\ifx\csname url\endcsname\relax
  \def\url#1{\texttt{#1}}\fi
\expandafter\ifx\csname urlprefix\endcsname\relax\def\urlprefix{URL }\fi
\providecommand{\bibinfo}[2]{#2}
\providecommand{\eprint}[2][]{\url{#2}}

\bibitem[{\citenamefont{Emary and Brandes}(2003)}]{Emary2003}
\bibinfo{author}{\bibfnamefont{C.}~\bibnamefont{Emary}} \bibnamefont{and}
  \bibinfo{author}{\bibfnamefont{T.}~\bibnamefont{Brandes}},
  \bibinfo{journal}{Physical Review E} \textbf{\bibinfo{volume}{67}},
  \bibinfo{pages}{066203} (\bibinfo{year}{2003}).

\bibitem[{\citenamefont{Chuchem et~al.}(2010)\citenamefont{Chuchem,
  Smith-Mannschott, Hiller, Kottos, Vardi, and Cohen}}]{Chuchem2010}
\bibinfo{author}{\bibfnamefont{M.}~\bibnamefont{Chuchem}},
  \bibinfo{author}{\bibfnamefont{K.}~\bibnamefont{Smith-Mannschott}},
  \bibinfo{author}{\bibfnamefont{M.}~\bibnamefont{Hiller}},
  \bibinfo{author}{\bibfnamefont{T.}~\bibnamefont{Kottos}},
  \bibinfo{author}{\bibfnamefont{A.}~\bibnamefont{Vardi}}, \bibnamefont{and}
  \bibinfo{author}{\bibfnamefont{D.}~\bibnamefont{Cohen}},
  \bibinfo{journal}{Phys. Rev. A} \textbf{\bibinfo{volume}{82}},
  \bibinfo{pages}{053617} (\bibinfo{year}{2010}).

\bibitem[{\citenamefont{Caprio et~al.}(2008)\citenamefont{Caprio, Cejnar, and
  Iachello}}]{Caprio2008}
\bibinfo{author}{\bibfnamefont{M.}~\bibnamefont{Caprio}},
  \bibinfo{author}{\bibfnamefont{P.}~\bibnamefont{Cejnar}}, \bibnamefont{and}
  \bibinfo{author}{\bibfnamefont{F.}~\bibnamefont{Iachello}},
  \bibinfo{journal}{Ann. Phys. (N.Y.)} \textbf{\bibinfo{volume}{323}},
  \bibinfo{pages}{1106 } (\bibinfo{year}{2008}).

\bibitem[{\citenamefont{Ribeiro et~al.}(2008)\citenamefont{Ribeiro, Vidal, and
  Mosseri}}]{Ribeiro2008}
\bibinfo{author}{\bibfnamefont{P.}~\bibnamefont{Ribeiro}},
  \bibinfo{author}{\bibfnamefont{J.}~\bibnamefont{Vidal}}, \bibnamefont{and}
  \bibinfo{author}{\bibfnamefont{R.}~\bibnamefont{Mosseri}},
  \bibinfo{journal}{Phys. Rev. E} \textbf{\bibinfo{volume}{78}},
  \bibinfo{pages}{021106} (\bibinfo{year}{2008}).

\bibitem[{\citenamefont{Brandes}(2013)}]{Brandes2013}
\bibinfo{author}{\bibfnamefont{T.}~\bibnamefont{Brandes}},
  \bibinfo{journal}{Phys. Rev. E} \textbf{\bibinfo{volume}{88}},
  \bibinfo{pages}{032133} (\bibinfo{year}{2013}).

\bibitem[{\citenamefont{Puebla et~al.}(2013)\citenamefont{Puebla, Rela\~no, and
  Retamosa}}]{Puebla2013}
\bibinfo{author}{\bibfnamefont{R.}~\bibnamefont{Puebla}},
  \bibinfo{author}{\bibfnamefont{A.}~\bibnamefont{Rela\~no}}, \bibnamefont{and}
  \bibinfo{author}{\bibfnamefont{J.}~\bibnamefont{Retamosa}},
  \bibinfo{journal}{Phys. Rev. A} \textbf{\bibinfo{volume}{87}},
  \bibinfo{pages}{023819} (\bibinfo{year}{2013}).

\bibitem[{\citenamefont{Scharf and Sundaram}(1992)}]{Scharf1992}
\bibinfo{author}{\bibfnamefont{R.}~\bibnamefont{Scharf}} \bibnamefont{and}
  \bibinfo{author}{\bibfnamefont{B.}~\bibnamefont{Sundaram}},
  \bibinfo{journal}{Phys. Rev. A} \textbf{\bibinfo{volume}{45}},
  \bibinfo{pages}{3615} (\bibinfo{year}{1992}).

\bibitem[{\citenamefont{Aubry et~al.}(1996)\citenamefont{Aubry, Flach, Kladko,
  and Olbrich}}]{Aubry1996}
\bibinfo{author}{\bibfnamefont{S.}~\bibnamefont{Aubry}},
  \bibinfo{author}{\bibfnamefont{S.}~\bibnamefont{Flach}},
  \bibinfo{author}{\bibfnamefont{K.}~\bibnamefont{Kladko}}, \bibnamefont{and}
  \bibinfo{author}{\bibfnamefont{E.}~\bibnamefont{Olbrich}},
  \bibinfo{journal}{Phys. Rev. Lett.} \textbf{\bibinfo{volume}{76}},
  \bibinfo{pages}{1607} (\bibinfo{year}{1996}).

\bibitem[{\citenamefont{Nissen and Keeling}(2010)}]{Nissen2010}
\bibinfo{author}{\bibfnamefont{F.}~\bibnamefont{Nissen}} \bibnamefont{and}
  \bibinfo{author}{\bibfnamefont{J.}~\bibnamefont{Keeling}},
  \bibinfo{journal}{Phys. Rev. A} \textbf{\bibinfo{volume}{81}},
  \bibinfo{pages}{063628} (\bibinfo{year}{2010}).

\bibitem[{\citenamefont{Caballero-Benítez
  et~al.}(2010)\citenamefont{Caballero-Benítez, Romero-Rochín, and
  Paredes}}]{Caballero-Benitez2010}
\bibinfo{author}{\bibfnamefont{S.~F.} \bibnamefont{Caballero-Benítez}},
  \bibinfo{author}{\bibfnamefont{V.}~\bibnamefont{Romero-Rochín}},
  \bibnamefont{and} \bibinfo{author}{\bibfnamefont{R.}~\bibnamefont{Paredes}},
  \bibinfo{journal}{J. Phys. B} \textbf{\bibinfo{volume}{43}},
  \bibinfo{pages}{115301} (\bibinfo{year}{2010}).

\bibitem[{\citenamefont{Juli\'a-D\'iaz
  et~al.}(2010{\natexlab{a}})\citenamefont{Juli\'a-D\'iaz, Martorell, and
  Polls}}]{Juli'a-D'iaz2010}
\bibinfo{author}{\bibfnamefont{B.}~\bibnamefont{Juli\'a-D\'iaz}},
  \bibinfo{author}{\bibfnamefont{J.}~\bibnamefont{Martorell}},
  \bibnamefont{and} \bibinfo{author}{\bibfnamefont{A.}~\bibnamefont{Polls}},
  \bibinfo{journal}{Phys. Rev. A} \textbf{\bibinfo{volume}{81}},
  \bibinfo{pages}{063625} (\bibinfo{year}{2010}{\natexlab{a}}).

\bibitem[{\citenamefont{Juli\'a-D\'iaz
  et~al.}(2010{\natexlab{b}})\citenamefont{Juli\'a-D\'iaz, Dagnino, Lewenstein,
  Martorell, and Polls}}]{Juli'a-D'iaz2010a}
\bibinfo{author}{\bibfnamefont{B.}~\bibnamefont{Juli\'a-D\'iaz}},
  \bibinfo{author}{\bibfnamefont{D.}~\bibnamefont{Dagnino}},
  \bibinfo{author}{\bibfnamefont{M.}~\bibnamefont{Lewenstein}},
  \bibinfo{author}{\bibfnamefont{J.}~\bibnamefont{Martorell}},
  \bibnamefont{and} \bibinfo{author}{\bibfnamefont{A.}~\bibnamefont{Polls}},
  \bibinfo{journal}{Phys. Rev. A} \textbf{\bibinfo{volume}{81}},
  \bibinfo{pages}{023615} (\bibinfo{year}{2010}{\natexlab{b}}).

\bibitem[{\citenamefont{Kawaguchi and Ueda}(2012)}]{Kawaguchi2012}
\bibinfo{author}{\bibfnamefont{Y.}~\bibnamefont{Kawaguchi}} \bibnamefont{and}
  \bibinfo{author}{\bibfnamefont{M.}~\bibnamefont{Ueda}},
  \bibinfo{journal}{Phys. Rep.} \textbf{\bibinfo{volume}{520}},
  \bibinfo{pages}{253 } (\bibinfo{year}{2012}).

\bibitem[{\citenamefont{Bernstein}(1993)}]{Bernstein1993}
\bibinfo{author}{\bibfnamefont{L.}~\bibnamefont{Bernstein}},
  \bibinfo{journal}{Phys. D (Amsterdam, Neth.)} \textbf{\bibinfo{volume}{68}},
  \bibinfo{pages}{174 } (\bibinfo{year}{1993}).

\bibitem[{\citenamefont{Franzosi et~al.}(2000)\citenamefont{Franzosi, Penna,
  and Zecchina}}]{Franzosi2000}
\bibinfo{author}{\bibfnamefont{R.}~\bibnamefont{Franzosi}},
  \bibinfo{author}{\bibfnamefont{V.}~\bibnamefont{Penna}}, \bibnamefont{and}
  \bibinfo{author}{\bibfnamefont{R.}~\bibnamefont{Zecchina}},
  \bibinfo{journal}{Int. J. Mod. Phys. B} \textbf{\bibinfo{volume}{14}},
  \bibinfo{pages}{943} (\bibinfo{year}{2000}).

\bibitem[{\citenamefont{Franzosi and Penna}(2001)}]{Franzosi2001}
\bibinfo{author}{\bibfnamefont{R.}~\bibnamefont{Franzosi}} \bibnamefont{and}
  \bibinfo{author}{\bibfnamefont{V.}~\bibnamefont{Penna}},
  \bibinfo{journal}{Phys. Rev. A} \textbf{\bibinfo{volume}{63}},
  \bibinfo{pages}{043609} (\bibinfo{year}{2001}).

\bibitem[{\citenamefont{Karkuszewski et~al.}(2002)\citenamefont{Karkuszewski,
  Sacha, and Smerzi}}]{Karkuszewski2002}
\bibinfo{author}{\bibfnamefont{Z.}~\bibnamefont{Karkuszewski}},
  \bibinfo{author}{\bibfnamefont{K.}~\bibnamefont{Sacha}}, \bibnamefont{and}
  \bibinfo{author}{\bibfnamefont{A.}~\bibnamefont{Smerzi}},
  \bibinfo{journal}{Eur. Phys. J. D} \textbf{\bibinfo{volume}{21}},
  \bibinfo{pages}{251} (\bibinfo{year}{2002}).

\bibitem[{\citenamefont{Buonsante et~al.}(2004)\citenamefont{Buonsante,
  Franzosi, and Penna}}]{Buonsante2004}
\bibinfo{author}{\bibfnamefont{P.}~\bibnamefont{Buonsante}},
  \bibinfo{author}{\bibfnamefont{R.}~\bibnamefont{Franzosi}}, \bibnamefont{and}
  \bibinfo{author}{\bibfnamefont{V.}~\bibnamefont{Penna}}, \bibinfo{journal}{J.
  Phys. B} \textbf{\bibinfo{volume}{37}}, \bibinfo{pages}{S229}
  (\bibinfo{year}{2004}).

\bibitem[{\citenamefont{Graefe and Korsch}(2007)}]{Graefe2007}
\bibinfo{author}{\bibfnamefont{E.~M.} \bibnamefont{Graefe}} \bibnamefont{and}
  \bibinfo{author}{\bibfnamefont{H.~J.} \bibnamefont{Korsch}},
  \bibinfo{journal}{Phys. Rev. A} \textbf{\bibinfo{volume}{76}},
  \bibinfo{pages}{032116} (\bibinfo{year}{2007}).

\bibitem[{\citenamefont{Juli\'a-D\'iaz
  et~al.}(2012{\natexlab{a}})\citenamefont{Juli\'a-D\'iaz, Torrontegui,
  Martorell, Muga, and Polls}}]{Juli'a-D'iaz2012}
\bibinfo{author}{\bibfnamefont{B.}~\bibnamefont{Juli\'a-D\'iaz}},
  \bibinfo{author}{\bibfnamefont{E.}~\bibnamefont{Torrontegui}},
  \bibinfo{author}{\bibfnamefont{J.}~\bibnamefont{Martorell}},
  \bibinfo{author}{\bibfnamefont{J.~G.} \bibnamefont{Muga}}, \bibnamefont{and}
  \bibinfo{author}{\bibfnamefont{A.}~\bibnamefont{Polls}},
  \bibinfo{journal}{Phys. Rev. A} \textbf{\bibinfo{volume}{86}},
  \bibinfo{pages}{063623} (\bibinfo{year}{2012}{\natexlab{a}}).

\bibitem[{\citenamefont{Juli\'a-D\'iaz
  et~al.}(2012{\natexlab{b}})\citenamefont{Juli\'a-D\'iaz, Zibold, Oberthaler,
  Mel\'e-Messeguer, Martorell, and Polls}}]{Juli'a-D'iaz2012a}
\bibinfo{author}{\bibfnamefont{B.}~\bibnamefont{Juli\'a-D\'iaz}},
  \bibinfo{author}{\bibfnamefont{T.}~\bibnamefont{Zibold}},
  \bibinfo{author}{\bibfnamefont{M.~K.} \bibnamefont{Oberthaler}},
  \bibinfo{author}{\bibfnamefont{M.}~\bibnamefont{Mel\'e-Messeguer}},
  \bibinfo{author}{\bibfnamefont{J.}~\bibnamefont{Martorell}},
  \bibnamefont{and} \bibinfo{author}{\bibfnamefont{A.}~\bibnamefont{Polls}},
  \bibinfo{journal}{Phys. Rev. A} \textbf{\bibinfo{volume}{86}},
  \bibinfo{pages}{023615} (\bibinfo{year}{2012}{\natexlab{b}}).

\bibitem[{\citenamefont{Hennig et~al.}(2012)\citenamefont{Hennig, Witthaut, and
  Campbell}}]{Hennig2012}
\bibinfo{author}{\bibfnamefont{H.}~\bibnamefont{Hennig}},
  \bibinfo{author}{\bibfnamefont{D.}~\bibnamefont{Witthaut}}, \bibnamefont{and}
  \bibinfo{author}{\bibfnamefont{D.~K.} \bibnamefont{Campbell}},
  \bibinfo{journal}{Phys. Rev. A} \textbf{\bibinfo{volume}{86}},
  \bibinfo{pages}{051604} (\bibinfo{year}{2012}).

\bibitem[{\citenamefont{P\'erez-Fern\'andez
  et~al.}(2011{\natexlab{a}})\citenamefont{P\'erez-Fern\'andez, Rela\~no,
  Arias, Cejnar, Dukelsky, and Garc\'ia-Ramos}}]{P'erez-Fern'andez2011a}
\bibinfo{author}{\bibfnamefont{P.}~\bibnamefont{P\'erez-Fern\'andez}},
  \bibinfo{author}{\bibfnamefont{A.}~\bibnamefont{Rela\~no}},
  \bibinfo{author}{\bibfnamefont{J.~M.} \bibnamefont{Arias}},
  \bibinfo{author}{\bibfnamefont{P.}~\bibnamefont{Cejnar}},
  \bibinfo{author}{\bibfnamefont{J.}~\bibnamefont{Dukelsky}}, \bibnamefont{and}
  \bibinfo{author}{\bibfnamefont{J.~E.} \bibnamefont{Garc\'ia-Ramos}},
  \bibinfo{journal}{Phys. Rev. E} \textbf{\bibinfo{volume}{83}},
  \bibinfo{pages}{046208} (\bibinfo{year}{2011}{\natexlab{a}}).

\bibitem[{\citenamefont{P\'erez-Fern\'andez
  et~al.}(2011{\natexlab{b}})\citenamefont{P\'erez-Fern\'andez, Cejnar, Arias,
  Dukelsky, Garc\'ia-Ramos, and Rela\~no}}]{P'erez-Fern'andez2011}
\bibinfo{author}{\bibfnamefont{P.}~\bibnamefont{P\'erez-Fern\'andez}},
  \bibinfo{author}{\bibfnamefont{P.}~\bibnamefont{Cejnar}},
  \bibinfo{author}{\bibfnamefont{J.~M.} \bibnamefont{Arias}},
  \bibinfo{author}{\bibfnamefont{J.}~\bibnamefont{Dukelsky}},
  \bibinfo{author}{\bibfnamefont{J.~E.} \bibnamefont{Garc\'ia-Ramos}},
  \bibnamefont{and} \bibinfo{author}{\bibfnamefont{A.}~\bibnamefont{Rela\~no}},
  \bibinfo{journal}{Phys. Rev. A} \textbf{\bibinfo{volume}{83}},
  \bibinfo{pages}{033802} (\bibinfo{year}{2011}{\natexlab{b}}).

\bibitem[{\citenamefont{P\'erez-Fern\'andez
  et~al.}(2009)\citenamefont{P\'erez-Fern\'andez, Rela\~no, Arias, Dukelsky,
  and Garc\'ia-Ramos}}]{P'erez-Fern'andez2009}
\bibinfo{author}{\bibfnamefont{P.}~\bibnamefont{P\'erez-Fern\'andez}},
  \bibinfo{author}{\bibfnamefont{A.}~\bibnamefont{Rela\~no}},
  \bibinfo{author}{\bibfnamefont{J.~M.} \bibnamefont{Arias}},
  \bibinfo{author}{\bibfnamefont{J.}~\bibnamefont{Dukelsky}}, \bibnamefont{and}
  \bibinfo{author}{\bibfnamefont{J.~E.} \bibnamefont{Garc\'ia-Ramos}},
  \bibinfo{journal}{Phys. Rev. A} \textbf{\bibinfo{volume}{80}},
  \bibinfo{pages}{032111} (\bibinfo{year}{2009}).

\bibitem[{\citenamefont{Bastidas et~al.}(2014)\citenamefont{Bastidas,
  P\'erez-Fern\'andez, Vogl, and Brandes}}]{Bastidas2014}
\bibinfo{author}{\bibfnamefont{V.~M.} \bibnamefont{Bastidas}},
  \bibinfo{author}{\bibfnamefont{P.}~\bibnamefont{P\'erez-Fern\'andez}},
  \bibinfo{author}{\bibfnamefont{M.}~\bibnamefont{Vogl}}, \bibnamefont{and}
  \bibinfo{author}{\bibfnamefont{T.}~\bibnamefont{Brandes}},
  \bibinfo{journal}{Phys. Rev. Lett.} \textbf{\bibinfo{volume}{112}},
  \bibinfo{pages}{140408} (\bibinfo{year}{2014}).

\bibitem[{\citenamefont{Winnewisser et~al.}(2005)\citenamefont{Winnewisser,
  Winnewisser, Medvedev, Behnke, De~Lucia, Ross, and Koput}}]{Winnewisser2005}
\bibinfo{author}{\bibfnamefont{B.~P.} \bibnamefont{Winnewisser}},
  \bibinfo{author}{\bibfnamefont{M.}~\bibnamefont{Winnewisser}},
  \bibinfo{author}{\bibfnamefont{I.~R.} \bibnamefont{Medvedev}},
  \bibinfo{author}{\bibfnamefont{M.}~\bibnamefont{Behnke}},
  \bibinfo{author}{\bibfnamefont{F.~C.} \bibnamefont{De~Lucia}},
  \bibinfo{author}{\bibfnamefont{S.~C.} \bibnamefont{Ross}}, \bibnamefont{and}
  \bibinfo{author}{\bibfnamefont{J.}~\bibnamefont{Koput}},
  \bibinfo{journal}{Phys. Rev. Lett.} \textbf{\bibinfo{volume}{95}},
  \bibinfo{pages}{243002} (\bibinfo{year}{2005}).

\bibitem[{\citenamefont{Zhao et~al.}(2014)\citenamefont{Zhao, Jiang, Tang,
  Webb, and Liu}}]{Zhao2014}
\bibinfo{author}{\bibfnamefont{L.}~\bibnamefont{Zhao}},
  \bibinfo{author}{\bibfnamefont{J.}~\bibnamefont{Jiang}},
  \bibinfo{author}{\bibfnamefont{T.}~\bibnamefont{Tang}},
  \bibinfo{author}{\bibfnamefont{M.}~\bibnamefont{Webb}}, \bibnamefont{and}
  \bibinfo{author}{\bibfnamefont{Y.}~\bibnamefont{Liu}},
  \bibinfo{journal}{Phys. Rev. A} \textbf{\bibinfo{volume}{89}},
  \bibinfo{pages}{023608} (\bibinfo{year}{2014}).

\bibitem[{\citenamefont{Dietz et~al.}(2013)\citenamefont{Dietz, Iachello,
  Miski-Oglu, Pietralla, Richter, von Smekal, and Wambach}}]{Dietz2013}
\bibinfo{author}{\bibfnamefont{B.}~\bibnamefont{Dietz}},
  \bibinfo{author}{\bibfnamefont{F.}~\bibnamefont{Iachello}},
  \bibinfo{author}{\bibfnamefont{M.}~\bibnamefont{Miski-Oglu}},
  \bibinfo{author}{\bibfnamefont{N.}~\bibnamefont{Pietralla}},
  \bibinfo{author}{\bibfnamefont{A.}~\bibnamefont{Richter}},
  \bibinfo{author}{\bibfnamefont{L.}~\bibnamefont{von Smekal}},
  \bibnamefont{and} \bibinfo{author}{\bibfnamefont{J.}~\bibnamefont{Wambach}},
  \bibinfo{journal}{Phys. Rev. B} \textbf{\bibinfo{volume}{88}},
  \bibinfo{pages}{104101} (\bibinfo{year}{2013}).

\bibitem[{\citenamefont{Zibold et~al.}(2010)\citenamefont{Zibold, Nicklas,
  Gross, and Oberthaler}}]{Zibold2010}
\bibinfo{author}{\bibfnamefont{T.}~\bibnamefont{Zibold}},
  \bibinfo{author}{\bibfnamefont{E.}~\bibnamefont{Nicklas}},
  \bibinfo{author}{\bibfnamefont{C.}~\bibnamefont{Gross}}, \bibnamefont{and}
  \bibinfo{author}{\bibfnamefont{M.~K.} \bibnamefont{Oberthaler}},
  \bibinfo{journal}{Phys. Rev. Lett.} \textbf{\bibinfo{volume}{105}},
  \bibinfo{pages}{204101} (\bibinfo{year}{2010}).

\bibitem[{\citenamefont{Gross et~al.}(2010)\citenamefont{Gross, Zibold,
  Nicklas, Est\`eve, and Oberthaler}}]{Gross2010}
\bibinfo{author}{\bibfnamefont{C.}~\bibnamefont{Gross}},
  \bibinfo{author}{\bibfnamefont{T.}~\bibnamefont{Zibold}},
  \bibinfo{author}{\bibfnamefont{E.}~\bibnamefont{Nicklas}},
  \bibinfo{author}{\bibfnamefont{J.}~\bibnamefont{Est\`eve}}, \bibnamefont{and}
  \bibinfo{author}{\bibfnamefont{M.~K.} \bibnamefont{Oberthaler}},
  \bibinfo{journal}{Nature (London)} \textbf{\bibinfo{volume}{464}},
  \bibinfo{pages}{1165} (\bibinfo{year}{2010}).

\bibitem[{\citenamefont{Steel and Collett}(1998)}]{Steel1998}
\bibinfo{author}{\bibfnamefont{M.~J.} \bibnamefont{Steel}} \bibnamefont{and}
  \bibinfo{author}{\bibfnamefont{M.~J.} \bibnamefont{Collett}},
  \bibinfo{journal}{Phys. Rev. A} \textbf{\bibinfo{volume}{57}},
  \bibinfo{pages}{2920} (\bibinfo{year}{1998}).

\bibitem[{\citenamefont{Baumann et~al.}(2010)\citenamefont{Baumann, Guerling,
  Brennecke, and Esslinger}}]{Baumann2010}
\bibinfo{author}{\bibfnamefont{K.}~\bibnamefont{Baumann}},
  \bibinfo{author}{\bibfnamefont{C.}~\bibnamefont{Guerling}},
  \bibinfo{author}{\bibfnamefont{F.}~\bibnamefont{Brennecke}},
  \bibnamefont{and}
  \bibinfo{author}{\bibfnamefont{T.}~\bibnamefont{Esslinger}},
  \bibinfo{journal}{Nature (London)} \textbf{\bibinfo{volume}{464}},
  \bibinfo{pages}{1301} (\bibinfo{year}{2010}).

\bibitem[{\citenamefont{Baumann et~al.}(2011)\citenamefont{Baumann, Mottl,
  Brennecke, and Esslinger}}]{Baumann2011}
\bibinfo{author}{\bibfnamefont{K.}~\bibnamefont{Baumann}},
  \bibinfo{author}{\bibfnamefont{R.}~\bibnamefont{Mottl}},
  \bibinfo{author}{\bibfnamefont{F.}~\bibnamefont{Brennecke}},
  \bibnamefont{and}
  \bibinfo{author}{\bibfnamefont{T.}~\bibnamefont{Esslinger}},
  \bibinfo{journal}{Phys. Rev. Lett.} \textbf{\bibinfo{volume}{107}},
  \bibinfo{pages}{140402} (\bibinfo{year}{2011}).

\bibitem[{\citenamefont{Baden et~al.}(2014)\citenamefont{Baden, Arnold,
  Grimsmo, Parkins, and Barrett}}]{Baden2014}
\bibinfo{author}{\bibfnamefont{M.~P.} \bibnamefont{Baden}},
  \bibinfo{author}{\bibfnamefont{K.~J.} \bibnamefont{Arnold}},
  \bibinfo{author}{\bibfnamefont{A.~L.} \bibnamefont{Grimsmo}},
  \bibinfo{author}{\bibfnamefont{S.}~\bibnamefont{Parkins}}, \bibnamefont{and}
  \bibinfo{author}{\bibfnamefont{M.~D.} \bibnamefont{Barrett}},
  \bibinfo{journal}{Phys. Rev. Lett.} \textbf{\bibinfo{volume}{113}},
  \bibinfo{pages}{020408} (\bibinfo{year}{2014}).

\bibitem[{\citenamefont{Paul and Uribe}(1993)}]{Paul1993}
\bibinfo{author}{\bibfnamefont{T.}~\bibnamefont{Paul}} \bibnamefont{and}
  \bibinfo{author}{\bibfnamefont{A.}~\bibnamefont{Uribe}},
  \bibinfo{journal}{Ann. I.H.P. Phys. Theor.} \textbf{\bibinfo{volume}{59}},
  \bibinfo{pages}{357} (\bibinfo{year}{1993}).

\bibitem[{\citenamefont{Lipkin et~al.}(1965)\citenamefont{Lipkin, Meshkov, and
  Glick}}]{Lipkin1965}
\bibinfo{author}{\bibfnamefont{H.}~\bibnamefont{Lipkin}},
  \bibinfo{author}{\bibfnamefont{N.}~\bibnamefont{Meshkov}}, \bibnamefont{and}
  \bibinfo{author}{\bibfnamefont{A.}~\bibnamefont{Glick}},
  \bibinfo{journal}{Nuclear Physics} \textbf{\bibinfo{volume}{62}},
  \bibinfo{pages}{188 } (\bibinfo{year}{1965}).

\bibitem[{\citenamefont{Meshkov et~al.}(1965)\citenamefont{Meshkov, Glick, and
  Lipkin}}]{Meshkov1965}
\bibinfo{author}{\bibfnamefont{N.}~\bibnamefont{Meshkov}},
  \bibinfo{author}{\bibfnamefont{A.}~\bibnamefont{Glick}}, \bibnamefont{and}
  \bibinfo{author}{\bibfnamefont{H.}~\bibnamefont{Lipkin}},
  \bibinfo{journal}{Nuclear Physics} \textbf{\bibinfo{volume}{62}},
  \bibinfo{pages}{199 } (\bibinfo{year}{1965}).

\bibitem[{\citenamefont{Glick et~al.}(1965)\citenamefont{Glick, Lipkin, and
  Meshkov}}]{Glick1965}
\bibinfo{author}{\bibfnamefont{A.}~\bibnamefont{Glick}},
  \bibinfo{author}{\bibfnamefont{H.}~\bibnamefont{Lipkin}}, \bibnamefont{and}
  \bibinfo{author}{\bibfnamefont{N.}~\bibnamefont{Meshkov}},
  \bibinfo{journal}{Nuclear Physics} \textbf{\bibinfo{volume}{62}},
  \bibinfo{pages}{211 } (\bibinfo{year}{1965}).

\bibitem[{\citenamefont{Dicke}(1954)}]{Dicke1954}
\bibinfo{author}{\bibfnamefont{R.~H.} \bibnamefont{Dicke}},
  \bibinfo{journal}{Phys. Rev.} \textbf{\bibinfo{volume}{93}},
  \bibinfo{pages}{99} (\bibinfo{year}{1954}).

\bibitem[{\citenamefont{Keeling}(2009)}]{Keeling2009}
\bibinfo{author}{\bibfnamefont{J.}~\bibnamefont{Keeling}},
  \bibinfo{journal}{Phys. Rev. A} \textbf{\bibinfo{volume}{79}},
  \bibinfo{pages}{053825} (\bibinfo{year}{2009}).

\bibitem[{\citenamefont{Tavis and Cummings}(1968)}]{Tavis1968}
\bibinfo{author}{\bibfnamefont{M.}~\bibnamefont{Tavis}} \bibnamefont{and}
  \bibinfo{author}{\bibfnamefont{F.~W.} \bibnamefont{Cummings}},
  \bibinfo{journal}{Phys. Rev.} \textbf{\bibinfo{volume}{170}},
  \bibinfo{pages}{379} (\bibinfo{year}{1968}).

\bibitem[{\citenamefont{Narducci et~al.}(1973)\citenamefont{Narducci, Orszag,
  and Tuft}}]{Narducci1973}
\bibinfo{author}{\bibfnamefont{L.~M.} \bibnamefont{Narducci}},
  \bibinfo{author}{\bibfnamefont{M.}~\bibnamefont{Orszag}}, \bibnamefont{and}
  \bibinfo{author}{\bibfnamefont{R.~A.} \bibnamefont{Tuft}},
  \bibinfo{journal}{Phys. Rev. A} \textbf{\bibinfo{volume}{8}},
  \bibinfo{pages}{1892} (\bibinfo{year}{1973}).

\bibitem[{\citenamefont{Dusuel and Vidal}(2005)}]{Dusuel2005}
\bibinfo{author}{\bibfnamefont{S.}~\bibnamefont{Dusuel}} \bibnamefont{and}
  \bibinfo{author}{\bibfnamefont{J.}~\bibnamefont{Vidal}},
  \bibinfo{journal}{Phys. Rev. B} \textbf{\bibinfo{volume}{71}},
  \bibinfo{pages}{224420} (\bibinfo{year}{2005}).

\bibitem[{\citenamefont{Casta\~nos et~al.}(2006)\citenamefont{Casta\~nos,
  L\'opez-Pe\~na, Hirsch, and L\'opez-Moreno}}]{Castanos2006}
\bibinfo{author}{\bibfnamefont{O.}~\bibnamefont{Casta\~nos}},
  \bibinfo{author}{\bibfnamefont{R.}~\bibnamefont{L\'opez-Pe\~na}},
  \bibinfo{author}{\bibfnamefont{J.~G.} \bibnamefont{Hirsch}},
  \bibnamefont{and}
  \bibinfo{author}{\bibfnamefont{E.}~\bibnamefont{L\'opez-Moreno}},
  \bibinfo{journal}{Phys. Rev. B} \textbf{\bibinfo{volume}{74}},
  \bibinfo{pages}{104118} (\bibinfo{year}{2006}).

\bibitem[{\citenamefont{Bastarrachea-Magnani
  et~al.}(2014)\citenamefont{Bastarrachea-Magnani, Lerma-Hern\'andez, and
  Hirsch}}]{Bastarrachea-Magnani2014a}
\bibinfo{author}{\bibfnamefont{M.~A.} \bibnamefont{Bastarrachea-Magnani}},
  \bibinfo{author}{\bibfnamefont{S.}~\bibnamefont{Lerma-Hern\'andez}},
  \bibnamefont{and} \bibinfo{author}{\bibfnamefont{J.~G.}
  \bibnamefont{Hirsch}}, \bibinfo{journal}{Phys. Rev. A}
  \textbf{\bibinfo{volume}{89}}, \bibinfo{pages}{032102}
  (\bibinfo{year}{2014}).

\bibitem[{\citenamefont{Ribeiro and Paul}(2009)}]{Ribeiro2009}
\bibinfo{author}{\bibfnamefont{P.}~\bibnamefont{Ribeiro}} \bibnamefont{and}
  \bibinfo{author}{\bibfnamefont{T.}~\bibnamefont{Paul}},
  \bibinfo{journal}{Phys. Rev. A} \textbf{\bibinfo{volume}{79}},
  \bibinfo{pages}{032107} (\bibinfo{year}{2009}).

\bibitem[{\citenamefont{Arecchi et~al.}(1972)\citenamefont{Arecchi, Courtens,
  Gilmore, and Thomas}}]{Arecchi1972}
\bibinfo{author}{\bibfnamefont{F.~T.} \bibnamefont{Arecchi}},
  \bibinfo{author}{\bibfnamefont{E.}~\bibnamefont{Courtens}},
  \bibinfo{author}{\bibfnamefont{R.}~\bibnamefont{Gilmore}}, \bibnamefont{and}
  \bibinfo{author}{\bibfnamefont{H.}~\bibnamefont{Thomas}},
  \bibinfo{journal}{Phys. Rev. A} \textbf{\bibinfo{volume}{6}},
  \bibinfo{pages}{2211} (\bibinfo{year}{1972}).

\bibitem[{\citenamefont{Zhang et~al.}(1990)\citenamefont{Zhang, Feng, and
  Gilmore}}]{Zhang1990}
\bibinfo{author}{\bibfnamefont{W.-M.} \bibnamefont{Zhang}},
  \bibinfo{author}{\bibfnamefont{D.~H.} \bibnamefont{Feng}}, \bibnamefont{and}
  \bibinfo{author}{\bibfnamefont{R.}~\bibnamefont{Gilmore}},
  \bibinfo{journal}{Rev. Mod. Phys.} \textbf{\bibinfo{volume}{62}},
  \bibinfo{pages}{867} (\bibinfo{year}{1990}).

\bibitem[{\citenamefont{Morrison and Parkins}(2008)}]{Morrison2008a}
\bibinfo{author}{\bibfnamefont{S.}~\bibnamefont{Morrison}} \bibnamefont{and}
  \bibinfo{author}{\bibfnamefont{A.~S.} \bibnamefont{Parkins}},
  \bibinfo{journal}{Phys. Rev. Lett.} \textbf{\bibinfo{volume}{100}},
  \bibinfo{pages}{040403} (\bibinfo{year}{2008}).

\bibitem[{\citenamefont{Kopylov et~al.}(2013)\citenamefont{Kopylov, Emary, and
  Brandes}}]{Kopylov2013}
\bibinfo{author}{\bibfnamefont{W.}~\bibnamefont{Kopylov}},
  \bibinfo{author}{\bibfnamefont{C.}~\bibnamefont{Emary}}, \bibnamefont{and}
  \bibinfo{author}{\bibfnamefont{T.}~\bibnamefont{Brandes}},
  \bibinfo{journal}{Phys. Rev. A} \textbf{\bibinfo{volume}{87}},
  \bibinfo{pages}{043840} (\bibinfo{year}{2013}).

\bibitem[{\citenamefont{Kopylov et~al.}(2015)\citenamefont{Kopylov, Emary,
  Sch\"{o}ll, and Brandes}}]{Kopylov2015}
\bibinfo{author}{\bibfnamefont{W.}~\bibnamefont{Kopylov}},
  \bibinfo{author}{\bibfnamefont{C.}~\bibnamefont{Emary}},
  \bibinfo{author}{\bibfnamefont{E.}~\bibnamefont{Sch\"{o}ll}},
  \bibnamefont{and} \bibinfo{author}{\bibfnamefont{T.}~\bibnamefont{Brandes}},
  \bibinfo{journal}{New J. Phys.} \textbf{\bibinfo{volume}{17}},
  \bibinfo{pages}{013040} (\bibinfo{year}{2015}).

\bibitem[{\citenamefont{Tonel et~al.}(2005)\citenamefont{Tonel, Links, and
  Foerster}}]{Tonel2005}
\bibinfo{author}{\bibfnamefont{A.~P.} \bibnamefont{Tonel}},
  \bibinfo{author}{\bibfnamefont{J.}~\bibnamefont{Links}}, \bibnamefont{and}
  \bibinfo{author}{\bibfnamefont{A.}~\bibnamefont{Foerster}},
  \bibinfo{journal}{J. Phys. A} \textbf{\bibinfo{volume}{38}},
  \bibinfo{pages}{1235} (\bibinfo{year}{2005}).

\bibitem[{\citenamefont{Gerving et~al.}(2012)\citenamefont{Gerving, Hoang,
  Land, Anquez, Hamley, and Chapman}}]{Gerving2012}
\bibinfo{author}{\bibfnamefont{C.}~\bibnamefont{Gerving}},
  \bibinfo{author}{\bibfnamefont{T.}~\bibnamefont{Hoang}},
  \bibinfo{author}{\bibfnamefont{B.}~\bibnamefont{Land}},
  \bibinfo{author}{\bibfnamefont{M.}~\bibnamefont{Anquez}},
  \bibinfo{author}{\bibfnamefont{C.}~\bibnamefont{Hamley}}, \bibnamefont{and}
  \bibinfo{author}{\bibfnamefont{M.}~\bibnamefont{Chapman}},
  \bibinfo{journal}{Nat. Commun.} \textbf{\bibinfo{volume}{3}},
  \bibinfo{pages}{1169} (\bibinfo{year}{2012}).

\bibitem[{\citenamefont{Buchleitner et~al.}(2002)\citenamefont{Buchleitner,
  Delande, and Zakrzewski}}]{Buchleitner2002}
\bibinfo{author}{\bibfnamefont{A.}~\bibnamefont{Buchleitner}},
  \bibinfo{author}{\bibfnamefont{D.}~\bibnamefont{Delande}}, \bibnamefont{and}
  \bibinfo{author}{\bibfnamefont{J.}~\bibnamefont{Zakrzewski}},
  \bibinfo{journal}{Phys. Rep.} \textbf{\bibinfo{volume}{368}},
  \bibinfo{pages}{409} (\bibinfo{year}{2002}).

\bibitem[{\citenamefont{Weaire and Srivastava}(1977)}]{Weaire1977}
\bibinfo{author}{\bibfnamefont{D.}~\bibnamefont{Weaire}} \bibnamefont{and}
  \bibinfo{author}{\bibfnamefont{V.}~\bibnamefont{Srivastava}},
  \bibinfo{journal}{J. Phys. C} \textbf{\bibinfo{volume}{10}},
  \bibinfo{pages}{4309} (\bibinfo{year}{1977}).

\bibitem[{\citenamefont{Khripkov et~al.}(2013)\citenamefont{Khripkov, Cohen,
  and Vardi}}]{Khripkov2013}
\bibinfo{author}{\bibfnamefont{C.}~\bibnamefont{Khripkov}},
  \bibinfo{author}{\bibfnamefont{D.}~\bibnamefont{Cohen}}, \bibnamefont{and}
  \bibinfo{author}{\bibfnamefont{A.}~\bibnamefont{Vardi}}, \bibinfo{journal}{J.
  Phys. A} \textbf{\bibinfo{volume}{46}}, \bibinfo{pages}{165304}
  (\bibinfo{year}{2013}).

\bibitem[{\citenamefont{Berry and Mount}(1972)}]{Berry1972}
\bibinfo{author}{\bibfnamefont{M.~V.} \bibnamefont{Berry}} \bibnamefont{and}
  \bibinfo{author}{\bibfnamefont{K.~E.} \bibnamefont{Mount}},
  \bibinfo{journal}{Rep. Prog. Phys.} \textbf{\bibinfo{volume}{35}},
  \bibinfo{pages}{315} (\bibinfo{year}{1972}).

\bibitem[{\citenamefont{Glauber}(1963)}]{Glauber1963}
\bibinfo{author}{\bibfnamefont{R.~J.} \bibnamefont{Glauber}},
  \bibinfo{journal}{Phys. Rev.} \textbf{\bibinfo{volume}{131}},
  \bibinfo{pages}{2766} (\bibinfo{year}{1963}).

\bibitem[{\citenamefont{{Oberthaler et al.}}()}]{Oberthaler}
\bibinfo{author}{\bibnamefont{{Oberthaler et al.}}}, \bibinfo{note}{(private
  communications)}.

\bibitem[{\citenamefont{Muessel et~al.}(2014)\citenamefont{Muessel, Strobel,
  Linnemann, Hume, and Oberthaler}}]{Muessel2014}
\bibinfo{author}{\bibfnamefont{W.}~\bibnamefont{Muessel}},
  \bibinfo{author}{\bibfnamefont{H.}~\bibnamefont{Strobel}},
  \bibinfo{author}{\bibfnamefont{D.}~\bibnamefont{Linnemann}},
  \bibinfo{author}{\bibfnamefont{D.~B.} \bibnamefont{Hume}}, \bibnamefont{and}
  \bibinfo{author}{\bibfnamefont{M.~K.} \bibnamefont{Oberthaler}},
  \bibinfo{journal}{Phys. Rev. Lett.} \textbf{\bibinfo{volume}{113}},
  \bibinfo{pages}{103004} (\bibinfo{year}{2014}).

\bibitem[{\citenamefont{Hamley et~al.}(2012)\citenamefont{Hamley, Gerving,
  Hoang, Bookjans, and Chapman}}]{Hamley2012}
\bibinfo{author}{\bibfnamefont{C.}~\bibnamefont{Hamley}},
  \bibinfo{author}{\bibfnamefont{C.}~\bibnamefont{Gerving}},
  \bibinfo{author}{\bibfnamefont{T.}~\bibnamefont{Hoang}},
  \bibinfo{author}{\bibfnamefont{E.}~\bibnamefont{Bookjans}}, \bibnamefont{and}
  \bibinfo{author}{\bibfnamefont{M.}~\bibnamefont{Chapman}},
  \bibinfo{journal}{Nat. Phys.} \textbf{\bibinfo{volume}{8}},
  \bibinfo{pages}{305} (\bibinfo{year}{2012}).

\bibitem[{\citenamefont{Bastidas et~al.}(2012)\citenamefont{Bastidas, Emary,
  Regler, and Brandes}}]{Bastidas2012b}
\bibinfo{author}{\bibfnamefont{V.~M.} \bibnamefont{Bastidas}},
  \bibinfo{author}{\bibfnamefont{C.}~\bibnamefont{Emary}},
  \bibinfo{author}{\bibfnamefont{B.}~\bibnamefont{Regler}}, \bibnamefont{and}
  \bibinfo{author}{\bibfnamefont{T.}~\bibnamefont{Brandes}},
  \bibinfo{journal}{Phys. Rev. Lett.} \textbf{\bibinfo{volume}{108}},
  \bibinfo{pages}{043003} (\bibinfo{year}{2012}).

\bibitem[{\citenamefont{Engelhardt et~al.}(2013)\citenamefont{Engelhardt,
  Bastidas, Emary, and Brandes}}]{Engelhardt2013}
\bibinfo{author}{\bibfnamefont{G.}~\bibnamefont{Engelhardt}},
  \bibinfo{author}{\bibfnamefont{V.~M.} \bibnamefont{Bastidas}},
  \bibinfo{author}{\bibfnamefont{C.}~\bibnamefont{Emary}}, \bibnamefont{and}
  \bibinfo{author}{\bibfnamefont{T.}~\bibnamefont{Brandes}},
  \bibinfo{journal}{Phys. Rev. E} \textbf{\bibinfo{volume}{87}},
  \bibinfo{pages}{052110} (\bibinfo{year}{2013}).

\end{thebibliography}

 \appendix
 
\section{Classical equation of motion for the Tavis-Cummings model}

A semiclassical investigation of the TC model has been  performed in Refs.~\cite{P'erez-Fern'andez2011,Bastarrachea-Magnani2014a}.
Here we  present a different approach to illustrate the semiclassical dynamics.

 The Heisenberg equations of motion of the operators in the TC model read
\begin{align}
 \frac d {dt} \hat a &= -i \omega   \hat a - i \frac \lambda {\sqrt N}  J_- \\
 \frac d {dt} J_x &= -\omega_0  J_y - \frac\lambda {\sqrt N}  \frac 1{i} \left( \hat a - \hat a^\dagger  \right)  J_z \\
 \frac d {dt} J_y &= \omega_0  J_x - \frac\lambda {\sqrt N}   \left( \hat a + \hat a^\dagger  \right)  J_z \\
 \frac d {dt} J_z &= -i \frac\lambda {\sqrt N} \left( \hat a  J_+ -  \hat a^\dagger  J_- \right).
 \label{eq:EOMxp}
\end{align}
In the thermodynamic limit $j \rightarrow \infty$, we consider the equations of motion for the classical variables $a e^{i \vartheta}=\hat a /\sqrt{j}$ and $j_{\alpha}=J_{\alpha} /j$,
with $a,\vartheta, j_{\alpha} \in \mathbb R$.
The equations of motion of these new coordinates then read
\begin{align}
\frac d {dt} a &=    - \frac \lambda {\sqrt{2}}  \ \left( \sin \vartheta j_x + \cos \vartheta j_y  \right)    \label{eq:EomTcTdlStart} \\
 \frac d {dt} \vartheta &= -\omega -\frac \lambda {\sqrt 2 a}   \left( \cos\vartheta j_x - \sin \vartheta j_y  \right) \label{eq:EomTcTdl2} \\
 \frac d {dt}  j_{x} &= -\omega_0  j_{y} - \lambda \sqrt{2} a   \sin \vartheta j_z\\
 \frac d {dt}   j_{y} &=  \omega_0   j_{x} -  \lambda \sqrt{2} a   \cos \vartheta j_z\\
 \frac d {dt}  j_{z}      & =  \lambda \sqrt{2}  a   \left( \sin \vartheta j_x + \cos \vartheta j_y  \right)
        \label{eq:EomTcTdlEnd}
 \end{align}
In  these new coordinates, the particle conservation in the thermodynamic limit is
\begin{equation}
 \mathcal M \equiv \lim_{j \rightarrow \infty} \frac {\hat M } j =  j_z + a^2.
 \label{eq:symmetryMTDL}
\end{equation}
Due to this conserved quantity, there is a constrain for the solutions $a$ and $j_z$. 
We define the rotated angular momentum $\tilde j_x = j_x \cos \vartheta - j_y \sin \vartheta$, 
$\tilde j_y = j_x \sin \vartheta + j_y \cos  \vartheta$ and $\tilde j_z = j_z$.
After some algebraic manipulations of Eqs. \eqref{eq:EomTcTdlStart}-\eqref{eq:EomTcTdlEnd}, we obtain the equations of motion for 
the rotated angular momentum.
By using the equations of motion for the classical variables $a, \vartheta$ describing the bosonic mode, Eqs. \eqref{eq:symmetryMTDL} and \eqref{eq:EomTcTdl2}, 
we obtain 
\begin{align}
  \frac d {dt} \tilde j_{x} &= -\omega_0  \tilde j_{y} -  \left( -\omega -\frac \lambda {\sqrt{2( \mathcal M- \tilde j_{z})}}  \tilde j_{x}  \right)  \tilde j_{y} \\
  \frac d {dt} \tilde j_{y} &=  \omega_0  \tilde{j}_{x} -  \lambda   \sqrt{2( \mathcal M-\tilde j_{z})}  \tilde j_{z}   \\  &+ \left(- \omega -\frac \lambda {\sqrt{2( \mathcal M- \tilde j_{z})}}  \tilde  j_{x}  \right)        \tilde  j_{ x} \\
 \frac d {dt} \tilde j_{z}      & = \lambda  \sqrt{2( \mathcal M- \tilde j_{z})}   \tilde j_{y}
\end{align}
These equations of motion can be directly derived starting from the effective energy landscape  
\begin{equation}
 E_{TC}\left( \tilde {\mathbf{j}}\right)=  \omega \left( \mathcal M - \tilde j_z \right) + \omega_0 \tilde j_z + \lambda \sqrt{2\left(M- \tilde j_z \right)} \tilde j_{ x},
\end{equation}
by evaluating the Poisson brackets, namely, $\frac d{dt} \tilde  j_\alpha  = \left\lbrace \tilde j_\alpha, E_{\text{TC}}\left( \tilde {\mathbf{j}}\right)   \right\rbrace$.
For this derivation one has to take into account the following 
Poisson bracket relations:
\begin{align}
 \left \lbrace \sqrt{2 \left( \mathcal M - \tilde{j}_z \right)}, \tilde{j}_x  \right \rbrace &=\frac{- \tilde{j}_y}{\sqrt{2 \left( \mathcal M - \tilde{j}_z \right)}} \\
 \left \lbrace \sqrt{2 \left( \mathcal M - \tilde{j}_z \right)}, \tilde{j}_y  \right \rbrace &=\frac{\tilde{j}_x}{\sqrt{2 \left( \mathcal M - \tilde{j}_z \right)}} .
\end{align}
These relations can be derived by representing the angular momentum operators with the Cartesian position and
 momentum operators.

\end{document}